\documentclass[a4paper,11pt]{article}
\pdfoutput=1 

\usepackage{jcappub} 
\usepackage[utf8]{inputenc}
\usepackage[numbers,sort&compress]{natbib}

\newcommand{\be}{\begin{equation}}
\newcommand{\ee}{\end{equation}}
\newcommand{\bea}{\begin{eqnarray}}
\newcommand{\eea}{\end{eqnarray}}

\newcommand{\abs}[1]{\lvert#1\rvert}

\DeclareMathAlphabet{\mathcal}{OMS}{cmsy}{m}{n}

\title{Stability of scalar perturbations in scalar-torsion $f(T,\phi)$ gravity theories in the presence of a matter fluid}

\author[a]{Manuel Gonzalez-Espinoza,}
\author[a]{Giovanni Otalora}
\author[a]{and \\ Joel Saavedra}

\affiliation[a]{Instituto de F\'{\i}sica, Pontificia Universidad Cat\'olica de Valpara\'{\i}so,\\ Casilla 4950,  Valpara\'{\i}so, Chile \label{addr1}}

\emailAdd{manuel.gonzalez@pucv.cl}
\emailAdd{giovanni.otalora@pucv.cl}
\emailAdd{\\ joel.saavedra@pucv.cl}

\abstract{We study the viability conditions for the absence of ghost, gradient and tachyonic instabilities, in scalar-torsion $f(T,\phi)$ gravity theories in the presence of a general barotropic perfect fluid. To describe the matter sector, we use the Sorkin-Schutz action and then calculate the second order action for scalar perturbations. For the study of ghost and gradient instabilities, we found that the gravity sector keeps decoupled from the matter sector and then applied the viability conditions for each one separately. Particularly, we verified that this theory is free from ghost and gradient instabilities, obtaining the standard results for matter, and for the gravity sector we checked that the corresponding speed of propagation satisfies $c_{s,g}^2=1$. On the other hand, in the case of tachyonic instability, we obtained the general expressions for the mass eigenvalues and then  evaluated them in the scaling matter fixed points of a concrete model of dark energy.  Thus, we found a space of parameters where it is possible to have a stable configuration respecting the constraints from the CMB measurements and the BBN constraints for early dark energy. Finally, we have numerically corroborated these results by solving the cosmological equations for a realistic cosmological evolution with phase space trajectories undergoing scaling matter regimes, and then showing that the system presents a stable configuration throughout cosmic evolution. }

\begin{document}
\maketitle
\flushbottom


\section{Introduction}
In 1998, the analysis of data of supernova Ia (SnIa) revealed that our Universe is expanding at an accelerating rate \cite{Riess:1998cb,Perlmutter:1998np}. But there is still no final interpretation that explains this fact. The more accepted interpretation is given by dark energy, a new form of exotic matter or modification to gravity, which is responsible for the accelerated expansion, and it constitutes 68 \% of the matter-energy density of the Universe \cite{Aghanim:2018eyx,Ade:2015rim}. And although the standard cosmology, based on Einstein's General Relativity, has obtained excellent results with the proposal that a cosmological constant $\Lambda$ is responsible for the accelerated expansion, this $\Lambda$CDM model (cosmological constant $\Lambda$ and cold dark matter) is plagued by a severe fine tuning problem associated with its energy scale \cite{Bull:2015stt,Martin:2012bt,Copeland:2006wr,amendola2010dark}. Moreover, some statistically-significant tensions with the latest data analysis have recently been detected when assuming the $\Lambda$CDM model. For instance, the $H_0$ disagreement between the CMB and the direct local distance ladder measurements \cite{Riess:2011yx,Riess:2016jrr,Riess:2018byc,DiValentino:2020zio}, the tension of the Planck data with weak lensing measurements and redshift surveys  related to the matter energy density $\Omega_{m}$, and the structure growth rate ($f\sigma 8$) \cite{Hildebrandt:2016iqg,Kuijken:2015vca,Conti:2016gav,DiValentino:2018gcu,DiValentino:2020vvd}. Although these tensions could mean a systematic bias, it is important to investigate the possibility of new physics beyond the standard cosmological model \cite{Riess:2019cxk, Davari:2019tni,DiValentino:2015bja,Sola:2019jek,Sola:2020lba,Joyce:2014kja,Koyama:2015vza}.

A viable and interesting alternative, widely studied in the literature, to explain dark energy consists in describing it in terms of a  scalar field such as in quintessence  \cite{Wetterich:1987fm,Ratra:1987rm,Carroll:1998zi,Tsujikawa:2013fta}, k-essence \cite{Chiba:1999ka,ArmendarizPicon:2000dh,ArmendarizPicon:2000ah}, tachyon fields \cite{Sen:2002nu,Sen:2002in}, amongst many others \cite{Copeland:2006wr,amendola2010dark}. From the viewpoint of quantum field theory in curved spacetime, a non-minimal coupling to gravity can naturally arise into the theory either by quantum corrections \cite{Linde:1982zj} or renormalizability requirements \cite{Freedman:1974gs,Freedman:1974ze,Birrell:1982ix}. So, for instance, a quintessence field coupled to gravity, the so-called `extended quintessence model' was firstly proposed in Ref. \cite{Perrotta:1999am}, and further studied in Refs. \cite{Sahni:1998at,Chiba:1999wt,Bartolo:1999sq,Faraoni:2000wk,Hrycyna:2008gk,Hrycyna:2007gd}. Also, a k-essence model with non-minimal coupling to gravity has been investigated in Ref. \cite{Sen:2008bg}, while a non-minimally coupled tachyonic field was considered in Ref. \cite{deSouza:2008nj}. In the context of Galileon models a non-minimal coupling to curvature allows to obtain second-order equations and thus avoiding pathological instabilities or the propagation of additional degrees of freedom \cite{Deffayet:2009wt}. Furthermore, it was recently shown that non-minimally coupled scalar field theories are capable of mitigating the current observational tensions in the concordance model \cite{Davari:2019tni,DiValentino:2019jae}.

Teleparallel Gravity (TG) is a gauge theory for the translation group, which introduces an equivalent description of gravity in terms of torsion \cite{Einstein,TranslationEinstein,Early-papers1,Early-papers2,Early-papers3,Early-papers4,Early-papers5,Early-papers6,JGPereira2,AndradeGuillenPereira-00,Arcos:2005ec,Pereira:2019woq}. The new dynamical variables are the tetrad fields replacing the usual metric tensor and the Weitzenb\"{o}ck connection substituting the usual Levi-Civita connection \cite{JGPereira2,AndradeGuillenPereira-00,Arcos:2005ec,Pereira:2019woq}. So, the Lagrangian density of TG is proportional to the torsion scalar $T$, which differs from the curvature scalar $R$ in a total derivative term. Therefore, the two theories are equivalent at the level of field equations \cite{Aldrovandi-Pereira-book,Arcos:2005ec}. In the same spirit of scalar-tensor theories, an immediately extension for TG is a non-minimally coupled scalar-torsion theory \cite{Cai:2015emx,Bahamonde:2017ize}. A scalar-torsion theory with non-minimal coupling term $\xi \phi^2 T$, where $\phi$ is the dynamical scalar field and $T$ the torsion scalar, with $\xi$ the coupling constant, was originally applied to dark energy in Ref. \cite{Geng:2011aj,Geng:2011ka}. Also, this theory was extended in Refs. \cite{Otalora:2013tba,Otalora:2013dsa} for both an arbitrary non-minimal coupling function $\phi^2\rightarrow F(\phi)$ and a tachyonic kinetic term for the scalar field. A key aspect of this extension is that, although TG coincides with GR at the level of field equations, a non-minimally coupled scalar-torsion theory is not equivalent to its counterpart based on curvature, that is to say, it belongs to a different class of gravitational modifications. 

A further extension of these theories can also be obtained when generalising the non-minimal coupling to gravity through introducing into the action terms in the form $F(\phi) G(T)$, where $G(T)$ is an arbitrary function of the torsion scalar $T$. This kind of gravitational modifications can also be included in a more general function $f(T,\phi)$, where it is also assumed an extension of the gravitational sector in analogy with $f(T)$ gravity \cite{Bengochea:2008gz,Linder:2010py,Li:2011wu}. In fact, an important generalisation of $f(T)$ gravity is obtained by allowing a general non-minimal coupling between the  torsion scalar and matter fields \cite{Harko:2014aja,Harko:2014sja,Carloni:2015lsa,Gonzalez-Espinoza:2018gyl}. This is in close analogy with the curvature-matter coupling in $f(R)$ gravity \cite{Nojiri:2004bi,Allemandi:2005qs,Nojiri:2006ri,Bertolami:2007gv,Harko:2008qz,Harko:2010mv,Bertolami:2009ic,Bertolami:2013kca,Wang:2013fja}, motivated by the counterterms appearing in the quantisation procedure of the self-interacting scalar field in curved spacetimes \cite{birrell1984quantum}. For instance, in the context of modified teleparallel gravity theories, and in order to explain the generation of primordial fluctuations during slow-roll inflation, the need of a generalised scalar-torsion $f(T, \phi)$ gravity theory was showed in Ref. \cite{Gonzalez-Espinoza:2020azh}. For late times, in the case of a concrete model of dark energy, the $f(T, \phi)$ gravity theories were investigated in Ref. \cite{Gonzalez-Espinoza:2020jss}, where the authors demonstrated the existence of new scaling solutions \cite{Uzan:1999ch,Amendola:1999qq} and attractors fixed points with accelerated expansion \cite{Copeland:2006wr,amendola2010dark}.

Even though these modified gravity theories are able to account for the observed accelerated expansion at late times and early inflation, they have at least an additional degree of freedom and it is important to ensure that the evolution of the associated modes does not produce pathological instabilities, e.g., ghost, Laplacian or tachyonic instabilities \cite{DeFelice:2016ucp,Heisenberg:2016eld,Kase:2014cwa,DeFelice:2011bh,Sbisa:2014pzo}. Furthermore, at the perturbation level, the modes related to the additional degrees of freedom are coupled to those associated with the degrees of freedom of the matter fields, and therefore, to perform a complete study of the stability conditions it is necessary to consider the interaction with matter \cite{Gergely:2014rna,Gleyzes:2014qga}. In this aspect, the Sorkin-Schutz action provides a general framework to describe the matter sector \cite{Schutz:1977df,Brown:1992kc}, allowing to expand the total action up to second order in perturbations to investigate the stability conditions in the presence of matter. This study is of vital importance to analyse the viability of the theory before comparing it with the full observational data \cite{Heisenberg:2016eld,DeFelice:2016ucp}. 

The manuscript is organised as follows: in Section \ref{fT_back_per}, we study the background equations of $f(T,\phi)$ and, by assuming the Sorkin-Schutz action to describe the matter sector, we study cosmological perturbations by expanding the action of $f(T,\phi)$ up to second order. In Section \ref{stability_conditions}, we calculate the stability conditions to avoid ghost, Laplacian and tachyonic instabilities. Finally, Section \ref{conclusion_f} is devoted to the conclusions.

\section{Scalar-Torsion $f(T,\phi)$ Gravity}\label{fT_back_per}

In the context of modified teleparallel gravity, and in the presence of matter, the action of the scalar-torsion $f(T,\phi)$ gravity theory is given by \cite{Gonzalez-Espinoza:2020jss}
\begin{equation}
 S=\int d^{4}x\,e\,\left[ f(T,\phi)+ P(\phi)X \right]+S_{m},
\label{action_Scalar_Torsion}
\end{equation} where $f(T,\phi)$ is an arbitrary function of the torsion scalar $T$ and the scalar field $\phi$. The kinetic term of the field is $X=-\partial_{\mu}{\phi}\partial^{\mu}{\phi}/2$ and $e\equiv \det(e^{A}_{~\mu})=\sqrt{-g}$. The torsion scalar $T$ is defined as 
\be
T= S_{\rho}^{~\mu\nu}\,T^{\rho}_{~\mu\nu},
\label{ScalarT}
 \ee
 where 
\bea \label{Def_Torsion}   
 && T^{\rho}_{~\mu\nu}\equiv e_{A}^{~\rho}\left[\partial_{\mu}e^{A}_{~\nu}
 -\partial_{\nu}e^{A}_{~\mu}+\omega^{A}_{~B\mu}\,e^{B}_{~\nu}
 -\omega^{A}_{~B\nu}\,e^{B}_{~\mu}\right],\\
&&  S_{\rho}^{~\mu\nu}=\frac{1}{2}\left(K^{\mu\nu}_{~~\rho}+\delta^{\mu}_{~\rho} \,T^{\theta\nu}_{~~\theta}-\delta^{\nu}_{~\rho}\,T^{\theta\mu}_{~~\theta}\right),\\ 
\label{Superpotential}
&& K^{\mu\nu}_{~~\rho}= -\frac{1}{2}\left(T^{\mu\nu}_{~~\rho}
 -T^{\nu\mu}_{~~\rho}-T_{\rho}^{~\mu\nu}\right),
\label{Contortion}
\eea are the components of the torsion tensor, the contortion tensor and the superpotential, respectively  \cite{Aldrovandi-Pereira-book,Arcos:2005ec}. The tetrad field $e^{A}_{~\mu}$, which locally relates the spacetime metric $g_{\mu\nu}$ and the Minkowski tangent space metric $\eta _{AB}^{}=\text{diag}\,(-1,1,1,1)$ through the local relation $g_{\mu\nu}=e^{A}_{~\mu}\,e^{B}_{~\nu}\,\eta_{AB}^{}$, plays the role of the dynamical variable of the theory. The non-trivial part of the tetrad field --the translational-valued gauge potential-- represents the gravitational field \cite{Arcos:2005ec,Pereira:2019woq}, while the inertial effects of the frame are stored in the purely inertial spin connection $\omega^{A}_{~B \mu}$ associated to the Weitzenb\"{o}ck connection \cite{Aldrovandi-Pereira-book,Pereira:2019woq}.

$S_{m}$ is the action of a general barotropic matter fluid which we define as the Sorkin-Schutz matter action. This latter reads as
\be
S_{m}=-\int{d^{4}x\left[e \rho(n)+J^{\nu}\partial_{\nu}\ell\right]},
\label{SS_action}
\ee where $\rho$ is the matter energy density which depends on the number density $n$, $\ell$ is a scalar field, and $J^{\nu}$ is a vector with weight one. 

Also, the number density $n$ is defined in the form
\be
n=\frac{\sqrt{-J^{\alpha}J^{\beta} g_{\alpha \beta}}}{e},
\ee and hence the four-velocity is expressed as
\be
u^{\alpha}=\frac{J^{\alpha}}{n e},
\label{u_Vel1}
\ee which satisfies the orthogonality relation $u^{\alpha}u_{\alpha}=-1$. 
The variation of the matter action \eqref{SS_action} with respect to $J^{\alpha}$ leads us to
\be
u_{\alpha}=\frac{1}{\rho_{,n}}\partial_{\alpha}{\ell}.
\label{u_Vel2}
\ee where we have defined $\rho_{,n}\equiv \partial{\rho}/\partial{n}$. On the other hand, the matter energy-momentum tensor is given by
\be
T_{\mu}^{~\nu}=e^{A}_{~\mu}\left[\frac{1}{e}\frac{\delta S_{m}}{\delta e^{A}_{~\nu}}\right]=n\rho_{,n} u_{\mu}u^{\nu}+\left(n\rho_{,n}-\rho\right)\delta^{\nu}_{\mu}, 
\ee which is the usual one associated with a perfect fluid such that
\be
p\equiv n \rho_{,n}-\rho.
\label{pressure}
\ee
So, as expected, for $\rho \propto n^{1+w}$ one gets  the relation $p=w \rho$. The conservation constraint is calculated by varying the matter action with respect to $\ell$, which yields
\be
\partial_{\alpha}{J^{\alpha}}=0.
\label{Conser_Part}
\ee 
Thus, varying the total action \eqref{action_Scalar_Torsion} with respect to the tetrad field $e^{A}_{~\mu}$ we obtain the field equations
\be
f_{,T} G_{\mu \nu}+S_{\mu \nu}{}^{\rho} \partial_{\rho} f_{,T}+\frac{1}{4}g_{\mu \nu}\left(f-T f_{,T}\right)+\frac{P}{4}\left(g_{\mu \nu} X+\partial_{\mu}\phi \partial_{\nu}\phi\right)=-\frac{1}{4}T_{\mu\nu},
\label{FieldEquations}
\ee where $G^{\mu}_{~\nu}=e_{A}^{~\mu} G^{A}_{~\nu}$ is the Einstein tensor, with $G_{A}^{~\mu}\equiv e^{-1}\partial_{\nu}\left(e e_{A}^{~\sigma} S_{\sigma}^{~\mu\nu}\right)-e_{A}^{~\sigma} T^{\lambda}_{~\rho \sigma}S_{\lambda}^{~\rho \mu}+e_{B}^{~\lambda} S_{\lambda}^{~\rho \mu}\omega^{B}_{~A \rho}+\frac{1}{4}e_{A}^{~\mu} T$ \cite{Aldrovandi-Pereira-book}. In the last step, the equation \eqref{FieldEquations} has been expressed in a general coordinate basis. In the second term of this equation there is an antisymmetric part associated with the tensor $S_{\mu \nu}{}^{\rho}$, which is consistent with the fact that the action \eqref{action_Scalar_Torsion} is not local Lorentz invariant \cite{Sotiriou:2010mv,Li:2010cg,Gonzalez-Espinoza:2020azh}. For TG, one has $f\sim T$, or equivalently $\partial_{\rho} f_{,T}=0$, and then the local Lorentz invariance is restored \cite{Aldrovandi-Pereira-book}. In the case of modified teleparallel gravity, the condition $\partial_{\rho} f_{,T} \neq 0$ is satisfied, and then, one is led to a set of six equations for six additional degrees of freedom, because of the violation of the local Lorentz symmetry.

\subsection{Background Equations}

At the cosmological setting, we choose the diagonal tetrad field
\begin{equation}
\label{veirbFRW}
e^A_{~\mu}={\rm
diag}(1,a,a,a),
\end{equation} which is a proper tetrad naturally associated with the vanishing spin connections $\omega^{A}_{~ B\mu}=0$ \cite{Krssak:2015oua}. This tetrad field leads to the flat Friedmann-Lema\^{i}tre-Robertson-Walker (FLRW) background
\begin{equation}
ds^2=-dt^2+a^2\,\delta_{ij} dx^i dx^j \,,
\label{FRWMetric}
\end{equation} where $a$ is the scale factor, function of the cosmic time $t$.

In this FLRW background, the fluid four-velocity in its rest frame is $u^{\mu}=(1,0,0,0)$, and using Eq. \eqref{u_Vel1} we get
\be
J^{0}=n a^3.
\ee  Thus, from Eq. \eqref{Conser_Part}, one obtains
\be
\mathcal{N}_{0}\equiv J^{0}=n a^3=constant,
\ee indicating that the particle number $\mathcal{N}_{0}$ is conserved. Furthermore, from this latter equation the number density $n$ satisfies the conservation equation 
\be
\dot{n}+3 H n=0,
\ee where $H=\dot{a}/a$, is the Hubble rate, a dot represents derivative with respect to $t$. Also, by using the relation \eqref{pressure} and the above equation, we obtain 
\be
\dot{\bar{\rho}}+3 H \left(\bar{\rho}+\bar{p}\right)=0.
\label{Contin_Eq}
\ee Here, we use the notation $\bar{\rho}$ and $\bar{p}$ for the energy and pressure densities of matter at the background, respectively, with equation of state $\bar{p}=w \bar{\rho}$, being $0 \leq w\leq 1$.

The background field equations calculated from the total action \eqref{action_Scalar_Torsion} are \cite{Gonzalez-Espinoza:2020jss}
\bea
\label{00}
 && f(T,\phi) - P(\phi) X - 2 T f_{,T}=\bar{\rho}, \\
\label{ii}
 && f(T,\phi) + P(\phi) X - 2 T f_{,T} - 4 \dot{H} f_{,T} - 4 H \dot{f}_{,T}=-\bar{p}, \\
\label{phi}
 &&  P(\phi) \ddot{\phi} + 3 P(\phi) H \dot{\phi} +P_{,\phi} X - f_{,\phi}=0,
\eea where a comma denotes the derivative with respect to $\phi$ or $T$, and the matter fluid satisfies the continuity equation \eqref{Contin_Eq}.

\subsection{Second order action}

To calculate the second order action, we use the Arnowitt-Deser-Misner (ADM) decomposition of the tetrad field \cite{Wu:2011kh}
\bea
&& e^{0}_{~\mu}=\left(N,\textbf{0}\right),\:\:\:\: e^{a}_{~\mu}=\left(N^{a},h^{a}_{~i}\right)\label{ADM1},\\
&& e_{0}^{~\mu}=\left(1/N,-N^{i}/N\right),\:\:\:\: e_{a}^{~\mu}=\left(0, h_{a}^{~i}\right)\label{ADM2},
\eea where $N^{i}=h_{a}^{~i}  N^{a}$, with $h^{a}_{~j} h_{a}^{~i}=\delta^{i}_{j}$, and $h^{a}_{~i}$ is the induced tetrad field.

In the uniform field gauge, $\delta \phi=0$, we take the ansatz
\be
N=1+\alpha,\:\:\:\: N^{a}=a^{-1} e^{-\mathcal{R}}\delta^{a}_{~i} \partial^{i}{\psi},\:\:\:\: h^{a}_{~i}=a e^{\mathcal{R}}\delta^{a}_{~j}\delta^{j}_{~i},
\label{Uniform_Field_Gauge}
\ee which leads us to the perturbed metric \cite{DeFelice:2011uc}
\be
 ds^2=-\left[\left(1+\alpha\right)^2-a^{-2}e^{-2\mathcal{R}}\left(\partial \psi\right)^2\right]dt^2 +2\partial_{i}{\psi}dt dx^{i}+ a^2 e^{2\mathcal{R}}\delta_{i j} dx^{i} dx^{j}.
\ee
Due to the violation of local Lorentz invariance in modified teleparallel gravity, we need to consider the corresponding additional degrees of freedom in the above perturbed tetrad field \cite{Izumi:2012qj,Golovnev:2018wbh}. Thus, we introduce them in the form of Goldstone modes of the symmetry breaking  \cite{Bluhm:2004ep, Bluhm:2007bd}, through the Lorentz rotation  
\be
\Lambda^{A}_{~B}=\left(e^{\chi}\right)^{A}_{~B}=\delta^{A}_{~B}+\chi^{A}_{~B}+\frac{1}{2} \chi^{A}_{~C} \chi^{C}_{~B}+\mathcal{O}(\chi^3),
\label{Lorentz_Transf}
\ee applied to the tetrad field, while keeping the vanishing spin connection of the background \cite{Wu:2016dkt,Gonzalez-Espinoza:2020azh}. Then the full perturbed tetrad field becomes
\be
 e'^{A}_{~\mu}=\left(e^{\chi}\right)^{A}_{~B} e^{B}_{~\mu}=e^{A}_{~\mu}+\chi^{A}_{~B}e^{B}_{~\mu}+\frac{1}{2} \chi^{A}_{~C} \chi^{C}_{~B} e^{B}_{~\mu}+\mathcal{O}(\chi^3),
\label{Transf_tetrad}
\ee
where the matrix $\chi_{A B}$ is antisymmetric and its components are given by 
\be
\chi^{0}_{~B}=\left(0,\chi_{b}\right),\:\:\:\: \chi^{a}_{~B}=\left(\chi^{a}, B^{a}_{~b}\right),
\ee with $\chi^{a}=\eta^{a b}\chi_{b}$ and  $B_{ab}=-B_{ba}$. Also, the spatial vector is defined by $\chi^{i}=h_{a}^{~i} \chi^{a}=\partial_{i}{\beta}+\chi^{(T)}_{i}$, and the spatial antisymmetric tensor is $B_{i j}=h^{a}_{~i} h^{b}_{~j} B_{a b}=-B_{j i}=-\epsilon_{j i k} B^{k}$. In general, the $\beta$ mode is scalar,  the $\chi^{(T)}_{i}$ mode is a transverse vector and the $B_{i}$ mode is a (pseudo) vector \cite{Wu:2016dkt,Golovnev:2018wbh,Gonzalez-Espinoza:2020azh}. But, in the present paper, we are only interested in the scalar mode $\beta$, and therefore, we neglect the other additional modes. 

Otherwise, we expand the fluid variables as follows
\bea
&& J^{0}=\mathcal{N}_{0}+\delta{J},\\
&& J^{i}=\frac{1}{a^2}\partial^{i}{\delta{j}},\\
&& \ell=-\ell_{0}-\rho_{,n} v,
\eea where $\partial_{0}\ell=-\rho_{,n} v$, and $v$ is the perturbation associated with the velocity potential \cite{Heisenberg:2016eld}. 

The matter energy density can be written as
\be
\rho=\bar{\rho}+\rho_{,n} \left[\frac{\delta J}{a^3}-\frac{3 \mathcal{N}_{0} \mathcal{R}}{a^3}\right]=\bar{\rho}\left(1+\delta_{M}\right),
\ee where we define the field variable $\delta_{M}\equiv \delta{\rho}/\bar{\rho}$, and hence we obtain
\be
\delta{J}=3 \mathcal{N}_{0} \mathcal{R}+\frac{a^3 \bar{\rho} \delta_{M}}{\rho_{,n}},
\ee which allows us to write $\delta{J}$ in terms of $\delta_{M}$ and $\mathcal{R}$. Also, from Eqs. \eqref{u_Vel1} and \eqref{u_Vel2} we find 
\be
\delta j=-\mathcal{N}_{0}\left(v+\psi-2 a^2 \beta\right).
\ee This latter equation is useful to eliminate $\delta{j}$ in favour of $v$, $\psi$ and $\beta$.

Thus, by putting all these pieces together, we expand the action \eqref{SS_action} up to second order in scalar perturbations to obtain 
\bea
&& S^{(2)}=\int dtd^{3}x a^{3}\Bigg\{\Big[- 2 (w+1)\partial^2 \beta+\frac{(w+1) \partial^2 \psi}{a^2}-\dot{\delta}_{M}- 3 (w+1)\dot{\mathcal{R}}\Big]v\bar{\rho}\nonumber\\
&& -\frac{(w+1)(\partial v)^2 \bar{\rho}}{2 a^2} + \Big[-8 F_1 H \partial^2 \beta+\frac{4 F_4 \partial^2 \mathcal{R} }{a^2}- 12 F_2 H \dot{\mathcal{R}}  \Big]\alpha-\Big[\frac{8 F_3 \partial^2 \beta }{3 a^2}-\frac{4 F_2 H \alpha }{a^2}\nonumber\\
&& +\frac{4 F_2 \dot{\mathcal{R}} }{a^2}\Big]\partial^2 \psi+\frac{4 F_3 (\partial^2 \psi)^2 }{3 a^4}-\frac{2 F_4 (\partial \mathcal{R})^2 }{a^2}+\frac{4 F_3 \left(\partial^2 \beta \right)^2 }{3}- \frac{w \delta_{M}^2 \bar{\rho}}{2 (w+1)}+\frac{F_6 \alpha^2}{2}\nonumber\\
&& -\alpha \delta_{M} \bar{\rho}+6 F_2 \dot{\mathcal{R}}^2 +8 F_1 \partial^2 \beta \dot{\mathcal{R}}+4 F_5 \partial^2 \beta \mathcal{R} \Bigg\},\label{SOAction}
\eea

where,
\begin{eqnarray}
F_1 &=& T f_{,TT}+f_{,T},\\
F_2 &=& 2T f_{,TT}+f_{,T},\\
F_3 &=& F_2 - F_1=T f_{,TT},\\
F_4 &=& 2 F_1 - F_2=f_{,T},\\
F_5 &=& \dot{F}_4, \\
F_6 &=& 2 T \left(2 T f_{,TT}+f_{,T}\right)+P(\phi) \dot{\phi}^2.
\end{eqnarray}

Below, we study the viability conditions for the absence of ghost, gradient and tachyonic instabilities.

\section{Stability Conditions}\label{stability_conditions}

Next, we use the Fourier transformation for the spatial coordinates, and write the second order action as
\bea
&& S^{(2)}=\frac{1}{(2\pi)^3}\int dtd^{3}k a^{3}\Bigg\{\Big[2 k^2 (w+1)\beta_{k}-\frac{k^2 (w+1) \psi_{k}}{a^2}-\dot{\delta}_{Mk}- 3 (w+1)\dot{\mathcal{R}}_{k}\Big]v_{k}\bar{\rho}\nonumber\\
&& -\frac{(w+1)(k v_{k})^2 \bar{\rho}}{2 a^2}+ \Big[8 F_1 k^2 H\beta_{k}-\frac{4 F_4 k^2 \mathcal{R}_{k} }{a^2}- 12 F_2 H \dot{\mathcal{R}}_{k}  \Big]\alpha_{k}+\Big[-\frac{8 F_3 k^4 \beta_{k} }{3 a^2}-\frac{4 F_2 k^2 H \alpha_{k} }{a^2}\nonumber\\
&& +\frac{4 F_2 k^2 \dot{\mathcal{R}}_{k} }{a^2}\Big]\psi_{k}+\frac{4 F_3 (k^2 \psi_{k})^2 }{3 a^4}-\frac{2 F_4 (k \mathcal{R}_{k})^2 }{a^2}+\frac{4 F_3 \left(k^2 \beta_{k}\right)^2 }{3}- \frac{w \delta_{Mk}^2 \bar{\rho}}{2 (w+1)}+\frac{F_6 \alpha_{k}^2}{2}\nonumber\\
&& -\alpha_{k} \delta_{Mk} \bar{\rho}+6 F_2 \dot{\mathcal{R}}_{k}^2 -8 F_1 k^2 \beta_{k} \dot{\mathcal{R}}_{k}-4 F_5 k^2 \beta_{k}\mathcal{R}_{k} \Bigg\}.\label{SOActionk}
\eea

Also, the above action contains two dynamical scalar modes $\{\mathcal{R}_{k},\delta_{Mk}\}$, and four auxiliary fields $\psi_{k}$, $\alpha_{k}$, $v_{k}$ and $\beta_{k}$. Then, by varying the action with respect to these auxiliary fields we obtain 
\bea
&& \bar{\rho}\left[-\frac{k^2}{a^2} \psi_{k}-3 \dot{\mathcal{R}}_{k}-\frac{1}{1+w}\dot{\delta}_{Mk}\right]+2 \bar{\rho} k^2 \beta_{k}-\bar{\rho} \frac{k^2}{a^2} v_{k}=0,\\
&& -4  F_4 \frac{k^2}{a^2} \mathcal{R}_{k}-12 F_2 H \dot{\mathcal{R}}_{k}+F_6 \alpha_{k} -
4 F_2 H \frac{k^2}{a^2}\psi_{k}-\delta_{Mk}\bar{\rho}+8 F_1 k^2 H \beta_{k} =0,\\
&& -2 F_5 \mathcal{R}_{k}-4 F_1 \dot{\mathcal{R}}_{k} + 4 F_1 H \alpha_{k}-\frac{4}{3} F_3 \frac{k^2}{a^2}\psi_{k}+\frac{4}{3} F_3 k^2 \beta_{k}+ (w+1) v_{k} \bar{\rho}=0,\\
&& (w+1) \bar{\rho} v_{k}+4 F_2 H \alpha_{k}-4 F_2 \dot{\mathcal{R}}_{k} -
\frac{8}{3} F_3 \frac{k^2}{a^2}\psi_{k}+\frac{8}{3} F_3 k^2 \beta_{k}=0.
\eea

By solving the above system of equations for the auxiliary fields, and after replacing these solutions into the action \eqref{SOActionk}, we get
\cite{DeFelice:2016ucp}
\be
S^{(2)}=\frac{1}{(2 \pi)^3}\int{d^{3}k dt a^3\left[\dot{\vec{\chi}}^{t}\mathbf{A}\dot{\vec{\chi}}-k^2\vec{\chi}^{t}\mathbf{G}\vec{\chi}-\dot{\vec{\chi}}^{t}\mathbf{B}\vec{\chi}-\vec{\chi}^{t}\mathbf{M}\vec{\chi}\right]},
\label{SOAction2}
\ee where the dimensionless vector $\vec{\chi}=\{\mathcal{R}_{k},\delta_{Mk}\}$ was defined, and the corresponding matrix components are shown in Appendix \ref{Appendix1}. 

By using the Sorkin-Schutz action to describe the matter sector, we were led to the second order action \eqref{SOAction} where there is not a quadratic kinetic term for the matter scalar mode. Then, it is not possible to directly elaborate for this action the usual Faddeev-Jackiw quantization procedure for constrained systems \cite{Faddeev:1988qp}. This same result was obtained in Refs. \cite{DeFelice:2016ucp,Kase:2019veo,Kase:2020hst}, in which the authors also used the Sorkin-Schutz action to describe matter and followed the same procedure we did to get an action similar to \eqref{SOAction2} in the unitary gauge. Nevertheless, it is possible to perform a transformation on the matter fields to overcome this problem. For instance, we can transform to the k-essence description of the perfect fluid \cite{DeFelice:2010gb,Arroja:2010wy,Giannakis:2005kr}, where the matter action has the appropriate form to carry out the Faddeev-Jackiw method \cite{Heisenberg:2016eld,Garriga:1997wz}. Although by following this path one should be led to an action different to \eqref{SOAction2} in the context of the k-essence description of matter, it is expected that the physical results derived from both actions have to be the same \cite{Heisenberg:2016eld}. We hope to return to this issue in a future work.

\subsection{Ghosts and Laplacian Instabilities}

As it is well known, a negative kinetic term in the action leads to a high energy vacuum, which is unstable to the spontaneous production of particles \cite{Carroll:2003st,Cline:2003gs}. Thus, in order to avoid this pathology the positivity of the kinetic term is demanded. Additionally, this constraint is imposed only in the high energy regime, as an infrared ghost does not imply a catastrophic vacuum collapse, but rather the Jeans instability \cite{Gumrukcuoglu:2016jbh}. Moreover, to study the Laplacian stability conditions, we also assume the high-$k$ limit when imposing the positivity of the propagation speed of the scalar modes.
This latter supposition is useful to evade complicated and non-local expressions because of the complex dependence on the momentum and the interaction between the fields in action \eqref{SOActionk}. This supposition is not only matter of simplicity because it is precisely in this regime that the gradient instability may acquire a high growth rate leading to a non-viable theory at 
the typical cosmological time scale \cite{DeFelice:2016ucp}. 

Thus, we proceed to diagonalise the kinetic matrix by taking 
\be
\mathcal{R}_{k}=\Psi_{1},\:\:\: \delta_{Mk}=k \Psi_{2}-\frac{A_{12} \Psi_{1}}{A_{22}},
\ee and then we obtain the Lagrangian density for second order perturbations
\be
\mathcal{L}^{(2)}=K_{11} \dot{\Psi}_{1}^2+K_{22} \dot{\Psi}_{2}^2+Q_{12} \left(\dot{\Psi}_{1} \Psi_{2}-\dot{\Psi}_{2} \Psi_{1}\right)-\mathcal{M}_{i j}\Psi_{i} \Psi_{j},
\label{L2_Cano_dia_I}
\ee where the coefficients $K_{11}$, $K_{22}$, $Q_{12}$ and $\mathcal{M}_{i,j}$ are shown in Appendix \ref{Appendix1}.

Particularly, in the high-$k$ limit the kinetic coefficients are written as
\be
K_{11}=A_{11}-\frac{A_{12}^2}{A_{22}}= \frac{P \dot{\phi}^2}{2 H^2},\:\:\:\: K_{22}=k^2 A_{22}= \frac{a^2 \bar{\rho}}{2 (w+1)}+\mathcal{O}(k^{-1}),
\ee and the second order action becomes
\bea
&& S^{(2)}=\frac{1}{(2\pi)^3}\int dtd^{3}k a^3\Bigg\{\frac{P \dot{\phi}^2}{2 H^2}\dot{\Psi}_{1}^2+\frac{a^2 \bar{\rho}}{2 (w+1)}\dot{\Psi}_{2}^2-\frac{k^2}{a^2}\left[\frac{P \dot{\phi}^2}{2 H^2}\Psi_{1}^2+\frac{w \bar{\rho} a^2}{2 (w+1)}\Psi_{2}^2\right]+\nonumber\\
&& \mathcal{O}(k^{-1})\Bigg\} .
\eea The conditions for the absence of ghosts instabilities reads $K_{11}>0$ and $K_{22}>0$, which therefore requires $P>0$ and $\bar{\rho}>0$.

Varying with respect to $\Psi_{1}$ and $\Psi_{2}$, we obtain
\bea
&& \ddot{\Psi}_{1}+\frac{k^2}{a^2}\Psi_{1}+\left[\frac{2 f_{,\phi}}{P \dot{\phi}}-\frac{(1+w) \bar{\rho}-4 H \dot{\phi} f_{,T\phi}+P \dot{\phi}^2}{24 H^3 f_{,TT}+2 H f_{,T}}-3 H\right]\dot{\Psi}_{1}\approx 0,\\
&& \ddot{\Psi}_{2}+ \frac{k^2}{a^2} w \Psi_{2}+(2-3 w) H \dot{\Psi}_{2} \approx 0.
\eea From these two latter equations we can read the propagation speed of the scalar modes for the gravity and matter sectors as
\be
c_{s,g}^2=1,\:\:\:\: c_{s}^2=w, 
\ee respectively. The results obtained in this section are consistent with what was found for the purely gravity sector in the vacuum in Ref. \cite{Gonzalez-Espinoza:2020azh}. 

\subsection{Tachyonic Instabilities}

Now, we investigate the canonical mass of the propagating modes and so the boundedness of the Hamiltonian at low momenta, which is related to the tachyonic instabilities and the expected Jeans instability in the matter sector \cite{Frusciante:2016xoj,DeFelice:2016ucp}.

Thus, following Ref. \cite{DeFelice:2016ucp}, we introduce the normalised fields 
\be
\Psi_{1}=\frac{\bar{\Psi}_{1}}{\sqrt{2 K_{11}}},\:\:\:\: \Psi_{2}=\frac{\bar{\Psi}_{2}}{\sqrt{2 K_{22}}},
\ee and after performing some integration by parts in \eqref{L2_Cano_dia_I}, we are led to the canonical form of the Lagrangian for the propagating modes 
\be
\mathcal{L}^{(2)}=\frac{a^{3}}{2}\left[\dot{\bar{\Psi}}_{1}^2+\dot{\bar{\Psi}}_{2}^2+\bar{B}(t,k)\left(\dot{\bar{\Psi}}_{1}\bar{\Psi}_{2}-\dot{\bar{\Psi}}_{2} \bar{\Psi}_{1}\right)-\bar{C}_{ij}(t,k)\bar{\Psi}_{i}\bar{\Psi}_{j}\right].
\label{L2_Cano}
\ee The expressions for $\bar{B}(t,k)$ and components $\bar{C}_{ij}(t,k)$ were put in Appendix \ref{Appendix1}.

Now, we diagonalise the matrix mass $\bar{C}_{ij}$ through the field rotation 
\be
\begin{bmatrix}
   \bar{\Psi}_{1} \\
   \bar{\Psi}_{2} 
\end{bmatrix} =\begin{bmatrix}
    \cos(\theta)& \sin(\theta) \\
    -\sin(\theta)& \cos(\theta)
\end{bmatrix} \begin{bmatrix}
    \Phi_{1} \\
    \Phi_{2} 
\end{bmatrix},
\ee where we also defined $\theta$ as
\be
\theta=-\frac{1}{2} \arctan{\left(\frac{2 \bar{C}_{12}}{\bar{C}_{11}-\bar{C}_{22}}\right)},
\ee 

Thus, the Lagrangian \eqref{L2_Cano} becomes \cite{DeFelice:2016ucp}
\be
\mathcal{L}^{(2)}=\frac{a^{3}}{2}\left[\dot{\Phi}_{1}^2+\dot{\Phi}_{2}^2+B(t,k)\left(\dot{\Phi}_{1}\Phi_{2}-\dot{\Phi}_{2} \Phi_{1}\right)-\mu_{1}(t,k)\Phi_{1}^2 -\mu_{2}(t,k)\Phi_{2}^2 \right],
\label{L2_Cano_dia}
\ee where 
\bea
\label{B}
&& B=\bar{B}+2 \dot{\theta},\\
\label{mu1}
&& \mu_{1}=-\dot{\theta}^2-\bar{B}\dot{\theta}+\frac{\left(\bar{C}_{11}-\bar{C}_{22}\right)^2+4 \bar{C}_{12}^2}{\bar{C}_{11}-\bar{C}_{22}} \cos^2(\theta)+\frac{\bar{C}_{11}\bar{C}_{22}-2\bar{C}_{12}^2-\bar{C}_{22}^2}{\bar{C}_{11}-\bar{C}_{22}},\\
&& \mu_{2}=-\dot{\theta}^2-\bar{B}\dot{\theta}-\frac{\left(\bar{C}_{11}-\bar{C}_{22}\right)^2+4 \bar{C}_{12}^2}{\bar{C}_{11}-\bar{C}_{22}} \cos^2(\theta)+\frac{\bar{C}_{11}^2-\bar{C}_{11}\bar{C}_{22}+2 \bar{C}_{12}^2}{\bar{C}_{11}-\bar{C}_{22}}.
\label{mu2}
\eea

For $k=0$, the tachyonic instability is avoided for $\mu_{i}(t,0)>0$, $i=1,2$. However, the eigenvalue $\mu_{2}$ may be negative as the dust sector will exhibit a Jeans instability, which is a necessary condition for structure formation. Thus, a theory is still viable for negative eigenvalues whenever they satisfy $\abs{\mu_{i}(t,0)}\lesssim H^2$ \cite{DeFelice:2016ucp}. 

\subsubsection{A concrete model}

In order to proceed forward we consider the class of models \cite{Gonzalez-Espinoza:2020azh,Gonzalez-Espinoza:2020jss}
\be
f(T,\phi)=-\frac{M_{pl}^2}{2} T-F(\phi)G(T)-V(\phi),
\label{Concre_model}
\ee where $F(\phi)$ is a non-minimal coupling function of $\phi$, and $V(\phi)$ is the scalar potential. 

Following Ref. \cite{Gonzalez-Espinoza:2020jss}, we obtain the modified Friedmann equations
\bea
\label{H00}
\frac{3}{\kappa^2} H^2 & = & G(T) F(\phi) - 2 T G_{,T} F(\phi) + V + P(\phi) X \nonumber\\
&& + \bar{\rho},\\
-\frac{2}{\kappa^2} \dot{H} & = & 2 P(\phi) X + 4 \dot{H} G_{,T} F(\phi) + 4 H G_{,TT} \dot{T} F(\phi) \nonumber\\
&& + 4 H G_{,T} \dot{F} + (1+w) \bar{\rho},
\label{Hii}
\eea
and the motion equation for $\phi$
\begin{equation}
P(\phi) \ddot{\phi} + 3 P(\phi) H \dot{\phi} + P_{,\phi} X + G(T) F_{,\phi}+ V_{,\phi}=0.
\end{equation}

These equations can be arranged in the following form  \cite{Copeland:2006wr}
\bea
\label{SH00}
&& \frac{3}{\kappa^2} H^2=\bar{\rho}_{de}+\bar{\rho},\\
&& -\frac{2}{\kappa^2} \dot{H}=\bar{\rho}_{de}+\bar{p}_{de}+(1+w) \bar{\rho},
\label{SHii}
\eea where the energy and pressure densities of dark energy are defined as
\bea
 \bar{\rho}_{de}&=& P(\phi) X + V - (2 T G_{,T} - G) F(\phi), \\
 \bar{p}_{de}&=&P(\phi) X - V + (2 T G_{,T} - G) F(\phi)\nonumber\\
&& + 4 ( 2 T G_{,TT} + G_{,T} ) F(\phi) \dot{H} + 4 H G_{,T} F_{,\phi}\dot{\phi},
\eea that obey the standard evolution equation 
\begin{eqnarray}
\dot{\bar{\rho}}_{de}+3H(1+w_{de})\bar{\rho}_{de}=0,
\end{eqnarray} in agreement with the energy conservation law and the fluid evolution equation
\bea
\label{rho}
 \dot{\bar{\rho}}+3 H (1+w)\bar{\rho}=0.
\label{rho_r}
\eea
We defined the effective dark energy equation-of-state parameter 
\begin{equation}
w_{de}=\frac{\bar{p}_{de}}{\bar{\rho}_{de}},
\label{wDE1}
\end{equation} and, as usual, it proves convenient to introduce the 
total equation-of-state parameter as 
\be
w_{tot}=\frac{\bar{p}_{de}+\bar{p}}{\bar{\rho}_{de}+\bar{\rho}},
\label{wtot}
\ee
which is immediately related to the deceleration parameter $q$ through 
\be
q=\frac{1}
{2}\left(1+3w_{tot}\right),
\label{deccelparam}
\ee
and hence the acceleration occurs when $q<0$. Also, we write the standard density 
parameters as 
\be
\Omega_{m}\equiv\frac{\kappa^2 \bar{\rho}}{3 H^2},\:\:\:\: \Omega_{de}\equiv\frac{\kappa^2 \bar{\rho}_{de}}{3 H^2},
\ee such that they satisfy
\be
\Omega_{de}+\Omega_{m}=1.
\ee

From now, we concentrate in the exponential potential $V(\phi)=V_{0} e^{-\lambda\kappa \phi}$, with $\lambda$ a dimensionless constant. Let us remember that this scalar potential can give rise to an accelerated expansion and to the same time allows to obtain cosmological scaling solutions \cite{amendola2010dark,Copeland:2006wr}. On the other hand, for the non-minimal coupling function of $\phi$ we take $F(\phi)=F_{0}e^{-\sigma \kappa \phi}$, such that $\sigma$ is a dimensionless constant. This is the most natural and simple choice for the non-minimal coupling function compatible with the exponential scalar potential \cite{Amendola:1999qq}. Also, we assume the ansatz  $G(T)=T^{1+s}/6^{1+s}$, and $P=1$. This ansatz can be seen as the immediate analogue of the non-linear matter-gravity coupling between curvature and a scalar field proposed in Refs. \cite{Nojiri:2004bi,Allemandi:2005qs} to explain dark energy and the cosmic acceleration. On the other hand, this non-linear coupling function in torsion is motivated from the physics of the very early universe and it is associated with the generation of primordial fluctuations during inflation in the context of $f(T,\phi)$ gravity \cite{Gonzalez-Espinoza:2020azh}. This non-linear coupling function in torsion was also studied in the context of dark energy in Ref. \cite{Gonzalez-Espinoza:2020jss}, where the authors found both new scaling solutions and new dark energy-dominated attractors. 

We introduce the following useful dimensionless variables \cite{Gonzalez-Espinoza:2020jss}
\be
x= \dfrac{\kappa \dot{\phi}}{\sqrt{6}H}, \:\:\: y= \dfrac{\kappa \sqrt{V}}{\sqrt{3} H}, \:\:\: u =-\frac{1}{3} (2 s+1) \kappa^2 F(\phi) H^{2 s}.
\label{Var}
\ee
In terms of these phase space variables the set of cosmological equations can be written as 
\bea
&& \frac{d x}{d N}=\frac{3 (w-1) x^3}{2 (s u+u-1)}+\frac{3 x \left[-2 s u+w \left(y^2+u-1\right)+y^2-u+1\right]}{2 (s u+u-1)}+\nonumber\\
&& \frac{\sqrt{\frac{3}{2}} \left[\lambda  (2 s+1) y^2 (s u+u-1)-\sigma  u \left(2 (s+1) x^2+s u+u-1\right)\right]}{(2 s+1) (s u+u-1)}\nonumber,\\
&& \frac{dy}{dN}=\frac{x y \left[3 (2 s+1) (w-1) x-\sqrt{6} (\lambda  (2 s+1) (s u+u-1)+2 (s+1) \sigma  u)\right]}{2 (2 s+1) (s u+u-1)}+\nonumber\\
&& \frac{3 (w+1) y \left(y^2+u-1\right)}{2 (s u+u-1)},\nonumber\\
&& \frac{du}{dN}=\frac{x u \left[-3 s (2 s+1) (w-1) x-\sqrt{6} \sigma  (s (u-2)+u-1)\right]}{(2 s+1) (s u+u-1)}-\nonumber\\
&& \frac{3 s (w+1) u \left(y^2+u-1\right)}{s u+u-1},
\label{Autonomous_Syst}
\eea where we also defined the e-folding number $N=\log(a)$. The critical points, their cosmological properties and stability conditions, were studied in Ref. \cite{Gonzalez-Espinoza:2020jss}. Particularly, the authors found new dark energy solutions which are attractors, and new scaling solutions which are saddle points representing the so-called scaling matter and radiation eras. 

\subsubsection{Scaling Regime}

The system satisfies the scaling solution 
\be
x=\frac{\sqrt{\frac{3}{2}} (w+1)}{\lambda },\:\:\: y=\sqrt{\frac{3}{2}} \sqrt{\frac{1-w^2}{\lambda ^2}},\:\:\: u=0.
\ee This critical point describes the scaling matter ($w=0$) and scaling radiation ($w=1/3$) epochs, that were denoted in Ref. \cite{Gonzalez-Espinoza:2020jss} by $g_{M}$ and $b_{R}$, respectively. For this point one has that $\Omega_{de}=3 (w+1)/\lambda ^2$ \cite{Gonzalez-Espinoza:2020jss}. This scaling solution is already present in the quintessence model \cite{Copeland:2006wr}, but the presence of a non-minimal coupling to torsion modifies its stability conditions in the phase space \cite{Gonzalez-Espinoza:2020jss}.

For this critical point, and $k=0$, we evaluate the mass eigenvalues \eqref{mu1} and \eqref{mu2} which gives
\bea
&& \frac{\mu_{1,2}}{H^2}=-\frac{27 w^3}{4 \lambda ^2}+\frac{3 \lambda ^2}{2 (w+1)}+\frac{9 \left(\lambda ^2+3\right) w}{4 \lambda ^2}+\frac{9 \left(\lambda ^2-3\right) w^2}{4 \lambda ^2}-\frac{3 \left(\lambda ^4+6 \lambda ^2-9\right)}{4 \lambda ^2}\pm\nonumber\\
&& \frac{3 (1-w) \left[\lambda ^2-3\left( w+1\right)\right] \sqrt{\left[ \lambda ^2-3 (w+1)^2\right]^2\left[\lambda ^4+18 \lambda ^2 (w+1)^2-27 (w+1)^4\right]}}{4 \lambda ^4 (w+1)}.
\eea
For $0 \leq w\leq 1$, and taking into account the reality condition $\Omega_{de}=3 (w+1)/\lambda ^2<1$, the eigenvalue $\mu_{1}$ is always positive or zero, while $\mu_{2}$ is always negative or zero. For $w=0$, the conditions for the absence of strong tachyonic instability are 
\be
|\lambda|>12.25,
\ee while for $w=1/3$ we obtain 
\be
|\lambda|>9.43.
\ee  We also considered the constraints for dark energy coming from the CMB measurements such that  $\Omega_{de}^{(m)}<0.02$ ($95\%$ C.L.) at redshift $z\approx 50$ \cite{Ade:2015rim}, and the Big Bang Nucleosynthesis (BBN) constraints for $\Omega_{de}^{(r)}<0.045$ \cite{Bean:2001wt}. 

The system has another scaling solution which exists only in the presence of a non-minimal coupling to torsion and $s\neq 0$. This new scaling solution is given by 
\be
x= -\frac{\sqrt{\frac{3}{2}} s (w+1)}{\sigma },\:\:\: y= 0,\:\:\: u=-\frac{3 s (2 s+1) \left(w^2-1\right)}{2 \sigma ^2}.
\ee This critical point describes the scaling matter ($w=0$) and scaling radiation ($w=1/3$) epochs, that were denoted in Ref. \cite{Gonzalez-Espinoza:2020jss} by $c_{R}$ and $i_{M}$, respectively, with $\Omega_{de}=3 s (w+1) (s (3-w)-w+1)/(2 \sigma^2)$.

For this critical point, and $k=0$, we evaluate the mass eigenvalues \eqref{mu1} and \eqref{mu2} which gives
\bea
&& \frac{\mu_{1,2}}{H^2}=\frac{9}{8} \left(w^2-9\right)-\frac{3 \left(\sigma ^2 w-\sigma ^2\right)}{4 s^2 (w+1)}+\frac{9 (w+1) \left(2 \sigma ^2+3 s w^2-3 s\right)}{2 \left(3 s^2 w^2-3 s^2+2 \sigma ^2+3 s w^2-3 s\right)}-\nonumber\\
&& \frac{9 \left(w^2-2 w+1\right)}{8 s}\pm\Bigg[\frac{3 \sigma ^2 (1-w)}{4 s^2 (w+1)}-\frac{9 (1-w) (3 s w+7 s-w+1)}{8 s}-\nonumber\\
&& \frac{81 s^4 (1-w) (w+1)^4}{\left[2 \sigma ^2+3 s (s+1) \left(w^2-1\right)\right]^2}+\frac{27 s^2 (1-w) (w+1)^2 (w+3)}{4 \sigma ^2+6 s (s+1) \left(w^2-1\right)}\Bigg]\times \nonumber\\
&& \sqrt{1+\frac{16 s^2 (w+1)^2 \left[6 \sigma ^2-9 s (w+1) (2 s-w+1)\right]}{\left[3 s (w+1) (s (w+3)-w+1)-2 \sigma ^2\right]^2}}.
\eea

\begin{figure}[htbp]
	\centering
		\includegraphics[width=0.6\textwidth]{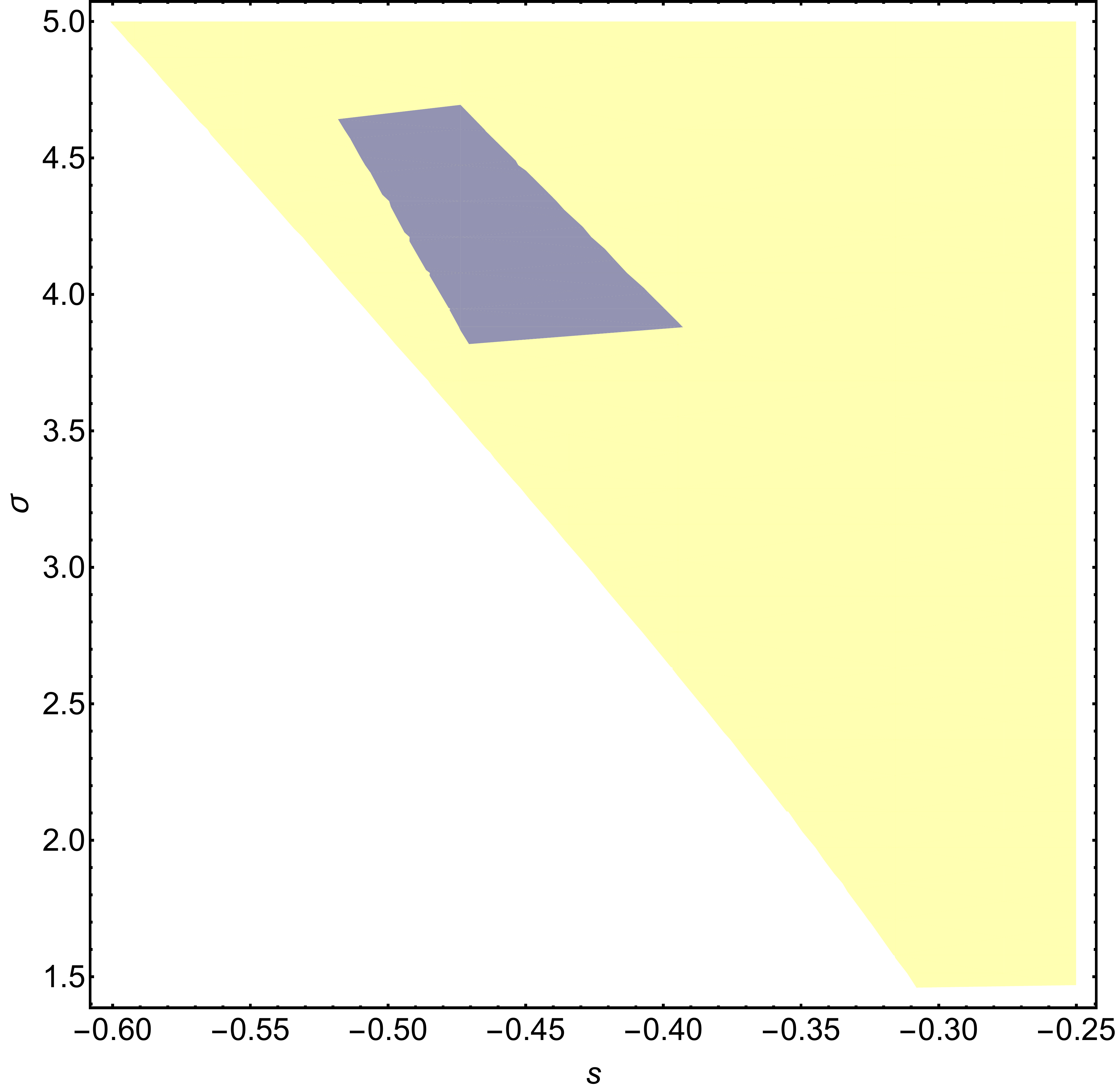}
	\caption{We plot the region of the space of parameters $\sigma$ and $s$ in which the conditions for the absence of rapidly evolving tachyonic instabilities are satisfied. The blue region corresponds to $w=0$, while the yellow one is associated with $w=1/3$. Outside these regions the theory can present a strong tachyonic instability during the scaling radiation and scaling matter eras, which would do it unviable. We also applied the constraints from the CMB measurements, such that $\Omega_{de}^{(m)}<0.02$ ($95\%$ C.L.) at redshift $z\approx 50$ \cite{Ade:2015rim}, and the BBN constraints for $\Omega_{de}^{(r)}<0.045$ \cite{Bean:2001wt}.}
	\label{Plot_Tachyonic_Inst_2}
\end{figure}

In FIG. \ref{Plot_Tachyonic_Inst_2}, we show the region of the space of parameters in which there are not strong tachyonic instabilities but rather the Jeans instability $|\mu_{2}|/H^2\lesssim 1$ that is essential for the formation of structure. As before, we also used the observational constraints for dark energy from the CMB measurements such that  $\Omega_{de}^{(m)}<0.02$ ($95\%$ C.L.) at redshift $z\approx 50$ \cite{Ade:2015rim}, and the Big Bang Nucleosynthesis (BBN) constraints for $\Omega_{de}^{(r)}<0.045$ \cite{Bean:2001wt}.

\subsubsection{Numerical Results}

In this section, we solve numerically the autonomous system \eqref{Autonomous_Syst} using the region of the space of parameters found in the above section for the scaling regimes. We also use the results obtained in Ref. \cite{Gonzalez-Espinoza:2020jss} for the stability of critical points and their cosmological properties. 

Thus, we solve for the two phase space trajectories $g_{M}\rightarrow k$ and $i_{M}\rightarrow h$, where $k$ and $h$ are attractors fixed points representing the dark energy dominated era. The critical point $h$ is given by 
\be
x=\frac{\lambda }{\sqrt{6}},\:\:\: y= \sqrt{1-\frac{\lambda ^2}{6}},\:\:\: u=0,
\ee which satisfies $\Omega_{de}=1$ and $w_{de}=w_{tot}=\frac{1}{3} \left(\lambda ^2-3\right)$. This point exists for $\abs{\lambda}<\sqrt{6}$ and it can explain the current cosmic acceleration for $\abs{\lambda}<\sqrt{2}$ \cite{Copeland:2006wr}. On the other hand, 
the critical point $k$ has the phase space components
\bea
&& x=\frac{-3 s (2 s+1)+\sqrt{A}}{\sqrt{6} (s+1) \sigma }, \:\:\: y=0,\:\:\: u=\frac{(2 s+1) \left[3 s \left(x_{c}^2+1\right)+\sqrt{6} \sigma  x_{c}\right]}{\sqrt{6} (s+1) \sigma  x_{c}+3 s (2 s+1)},
\eea with $A=9 s^2 (2 s+1)^2-6 s (s+1) \sigma ^2$. This point is a new dark energy dominated solution for $s\neq 0$, with $\Omega_{de}=1$, and $w_{de}=w_{tot}=\frac{3 s^2-\sqrt{A}}{3 s (s+1)}$. The conditions for existence and accelerated expansion were found in Ref. \cite{Gonzalez-Espinoza:2020jss}.

\begin{figure}[htbp]
	\centering
		\includegraphics[width=0.6\textwidth]{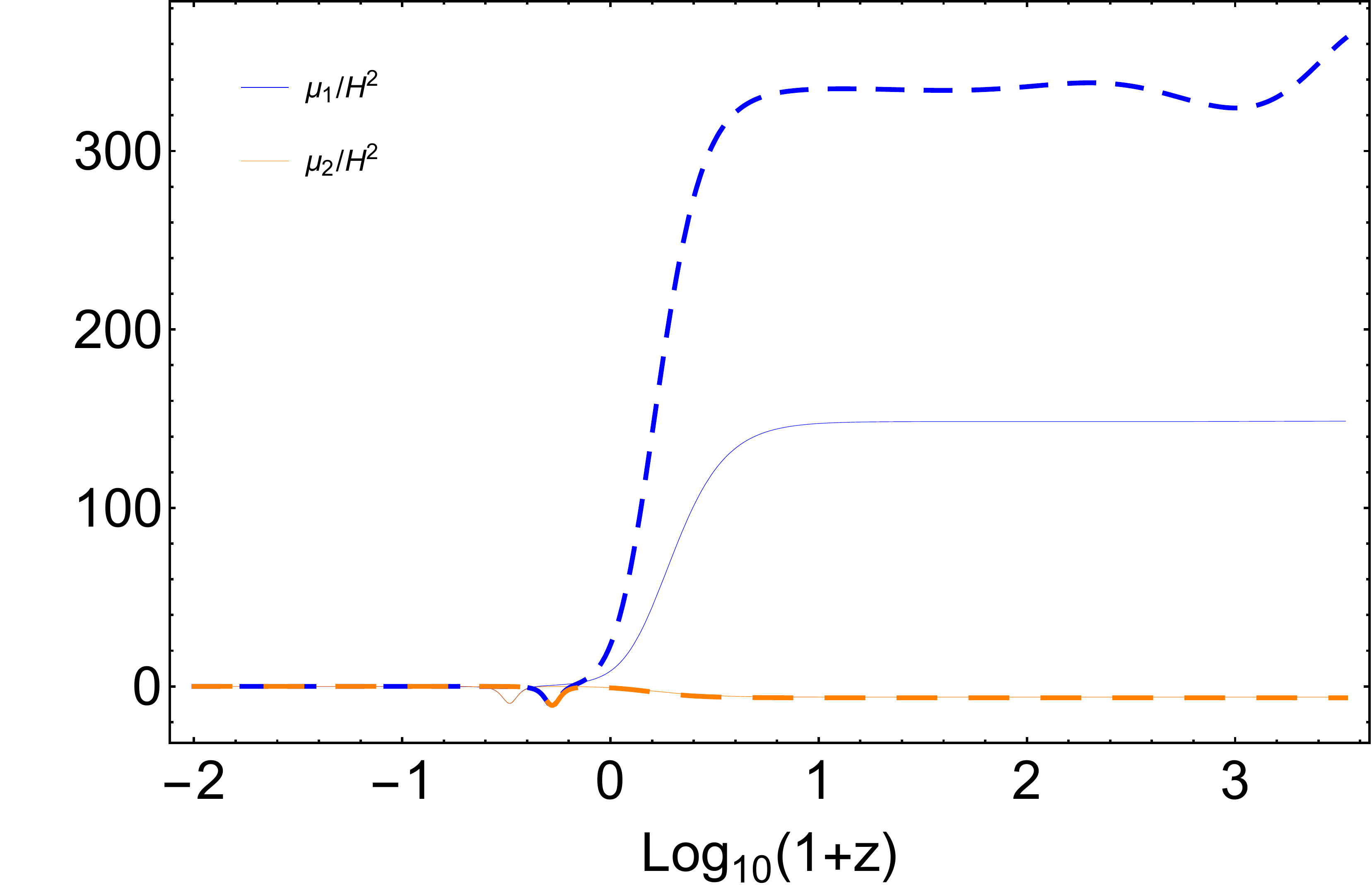}
	\caption{We depict the behaviour of the eigenvalues $\mu_{1}/H^2$ (blue line) and $\mu_{2}/H^2$ (orange line) at late times for the phase space trajectories $g_{M}\rightarrow k$ (dashed line) and $i_{M}\rightarrow h$ (solid line). In the first case we take the values $s=-1.20$, $\sigma=0.3$, $\lambda=15$, and initial conditions $x_{i}=0.08$, $y_{i}=0.08$ and $u_{i}=8.3\times 10^{-13}$. At the present time $z=0$ we get $\Omega_{de}^{(0)}\approx 0.68$, $\Omega_{m}^{(0)}\approx 0.32$ and $w_{de}^{(0)}\approx -1.033$. In the second case, we consider $s=-0.47$, $\sigma=4.68$, and $\lambda=0.3$, with initial conditions $x_{i}=0.123$, $y_{i}=1.04621 \times 10^{-5}$ and $u_{i}=-1.93 \times 10^{-3}$, to obtain at the present time $z=0$ the values $\Omega_{de}^{(0)}\approx 0.68$, $\Omega_{m}^{(0)}\approx 0.32$, and $w_{de}^{(0)}\approx -0.961$. In both cases, the results obtained are in agreement with the Planck data 2018 \cite{Aghanim:2018eyx}.}
	\label{Plot_Evol_2_3}
\end{figure}

In FIG. \ref{Plot_Evol_2_3} we show the evolution of the mass eigenvalues $\mu_{1}$ and $\mu_{2}$, Eqs. \eqref{mu1} and \eqref{mu2}, respectively, for the two phase space trajectories  $g_{M}\rightarrow k$ and $i_{M}\rightarrow h$. We use the values of the parameters found in the previous section. For the trajectory $g_{M}\rightarrow k$, corresponding to the solid blue and orange lines, we choose $s=-1.20$, $\sigma=0.3$ and $\lambda=15$. In this case, at redshift $z=0$, we obtain  $\Omega_{de}^{(0)}\approx 0.68$, $\Omega_{m}^{(0)}\approx 0.32$ and $w_{de}^{(0)}\approx-1.033$. Thus, it is observed that the system presents a tachyonically stable configuration. For the trajectory $i_{M}\rightarrow h$, corresponding in this case to the dashed blue and orange lines, we have $s=-0.47$, $\sigma=4.68$, and $\lambda=0.3$. At  redshift  $z=0$, we get the values $\Omega_{de}^{(0)}\approx 0.68$, $\Omega_{m}^{(0)}\approx 0.32$, and $w_{de}^{(0)}\approx -0.961$. In both cases, the results obtained are in agreement with the Planck data 2018 \cite{Aghanim:2018eyx}, and also, they lead us to a tachyonically stable configuration.

 \section{Concluding Remarks}\label{conclusion_f}

In the present paper, we investigated the stability of scalar perturbations in scalar-torsion $f(T,\phi)$ gravity theories \cite{Gonzalez-Espinoza:2020azh,Gonzalez-Espinoza:2020jss}, in the presence of a barotropic perfect fluid. We employ the Sorkin-Schutz action to describe the matter sector, and after calculating the second order action for scalar perturbations, we extracted the viability conditions for the absence of ghost, Laplacian and tachyonic instabilities.

Longe-range forces interactions can arise in the context of several different high-energy theories. Among these we have for example the extended supergravity models \cite{Scherk:1980gq,Sakstein:2017xjx}, the compactification of higher dimensional theories \cite{Bars:1986gt}, and the extensions of the standard electroweak model \cite{Hill:1988bu}. Although a long-range force of gravitational strength coupled to baryonic matter is in disagreement with the local experimental constraints \cite{Fackler:1989nv,Damour:1992kf,Will:2014kxa,Ishak:2018his,Burrage:2017qrf,LIGOScientific:2016aoc,LIGOScientific:2017ync}, in some extensions of the standard electroweak model, the ultralight pseudo-Nambu-Goldstone bosons with scalar couplings are propagating physical degrees of freedom which can mediate additional galactic-range forces in the dark matter sector \cite{Frieman:1991zxc}. One specific example of this class of models involves an ultra-light scalar field associated with the spontaneous breaking of some global symmetry that interacts with fermionic dark matter through a Yukawa coupling, and therefore, leading to a long-range `fifth' force in addition to gravity \cite{Gradwohl:1992ue}. However, these longe-range forces are strongly constrained from galaxy and cluster dynamics \cite{Gubser:2004du,Gubser:2004uh,Nusser:2004qu,Kesden:2006zb,Kesden:2006vz,Farrar:2006tb}, as well as in cosmological scales, through the large scale structure and cosmic microwave background data \cite{Sealfon:2004gz,Bean:2008ac,Bai:2015vca,Esteban:2021ozz}. On the other hand, additional degrees of freedom are found when studying the dynamics of cosmological perturbations in the context of modified teleparallel gravity due to the violation of local Lorentz symmetry.
These Goldstone modes associated with local Lorentz rotations of the tetrad field turn out to be fields that do not propagate as it can be observed in the second order action \eqref{SOAction}. The additional scalar mode $\beta$  behaves as an auxiliary field that does not propagate. Moreover, for tensor perturbations there are not additional degrees of freedom in the corresponding second order action \cite{Gonzalez-Espinoza:2019ajd,Gonzalez-Espinoza:2020azh}. There are only the usual transverse massless graviton modes propagating at the speed of light which is consistent with observations \cite{Baker:2017hug,Sakstein:2017xjx}. Finally, related to the local experiments, these generalised scalar-torsion $f(T,\phi)$ gravity theories without derivative couplings are indistinguishable from GR by their parametrised post-Newtonian (PPN) parameters \cite{Flathmann:2019khc,Chen:2014qsa,Li:2013oef}. 

We assumed the high-$k$ limit when deriving the conditions for no-ghost and no-gradient instabilities  \cite{Gumrukcuoglu:2016jbh}. As it is well known, only high-$k$ contributions to the Lagrangian can lead to catastrophic instabilities whereas other terms can be recast in mass-like terms through appropriate field redefinitions \cite{ DeFelice:2016ucp}. So, these conditions were obtained by imposing the positivity of the leading terms of the components of the kinetic matrix in the high-$k$ regime. Also, under the same approximation, we obtained the speed of propagation for the curvature and matter modes, verifying that the speed of propagation associated to the gravity sector does not depend on the fluid variables, and the standard expressions were obtained for the matter sector. These results are consistent with what was found in Ref. \cite{Gonzalez-Espinoza:2020azh}, where $c_{s,g}^2=1$ was obtained for the pure gravity sector in the vacuum.

Also, we investigated the conditions for the absence of tachyonic instabilities which appear when the Hamiltonian is unbounded from below at low energy. These instabilities are related to the presence of tachyonic mass terms in the second order action. After performing a transformation of the fields to write the Lagrangian density in its canonical form, and then by diagonalising the emergent mass matrix, we calculated the general expressions for the mass eigenvalues to be constrained in order to find the corresponding stability conditions. Thus, at the limit $k\rightarrow 0$, we evaluated these expressions in the scaling fixed points of the model defined in Eq. \eqref{Concre_model}. The cosmological dynamics of this model was studied in Ref. \cite{Gonzalez-Espinoza:2020jss}, where the authors showed that it can explain the current accelerated expansion, with new scaling solutions representing the so-called scaling radiation and scaling matter eras \cite{Albuquerque:2018ymr}. Besides their interesting cosmological properties \cite{Uzan:1999ch,Amendola:1999qq,Amendola:2006qi,Gomes:2013ema}, scaling solutions provide the simplest way to obtain analytical results for the different cosmological scenarios that include the effects of dark energy during the matter and radiation dominated eras \cite{amendola2010dark}. Therefore, using the analytical results valid for the scaling regimes, we constrained the free parameters of the model, and so, we found a region of the space of parameters where it is possible to have a tachyonically stable configuration, and for the which it is satisfied the observational constraints from the CBM measurements  \cite{Ade:2015rim} and the BBN constraints for early dark energy \cite{Bean:2001wt}.

Finally, we corroborated our analytical results by solving numerically the cosmological equations through an associated autonomous system for the corresponding phase space variables. By choosing the appropriate initial conditions to obtain a realistic evolution, with phase space trajectories undergoing scaling regimes, and adjusting the current values for the energy density parameter and the equation of state of dark energy from Planck \cite{Aghanim:2018eyx}, we verified that the system shows a tachyonically stable configuration throughout the cosmic evolution. In fact, the mass eigenvalue $\mu_{1}$ associated with the gravity sector is always positive, whereas the other one related to the matter sector is always negative and it satisfies $|\mu_{2}|/H^2\lesssim 1$, which is a clear manifestation of the Jeans instability, essential for the large-scale structure formation in the  Universe \cite{DeFelice:2016ucp}.

In the present paper we studied the stability of scalar perturbations in the generalised scalar-torsion $f(T,\phi)$ gravity in the presence of matter fields. The function $f(T,\phi)$ can include a scalar field which is non-minimally coupled to torsion, while the matter sector is always minimally coupled to gravity without any explicit dependence on the scalar field. This setup is similar to that of the scalar-tensor gravity theories in the Jordan frame, where the matter fields are universally coupled to the metric tensor in order to comply the weak equivalence principle (WEP) \cite{Tino:2020nla}. Nevertheless, a coupling between the scalar field and matter is also expected unless some unknown symmetry prevents or suppresses it \cite{Carroll:1998zi}. Also, the non-minimal coupling of the scalar field to gravity is related by a conformal transformation with the explicit interaction between the scalar field and matter \cite{Wetterich:1994bg,Amendola:1999qq}. In this way, a cosmological model in the context of a generalised Jordan-Brans-Dicke theory in which the scalar field couples with different strengths to baryonic matter and to dark matter was proposed in Ref. \cite{Damour:1990tw}. On the other hand, the study of the cosmological dynamics and the evolution of the cosmological perturbations for the quintessence scalar field with an universal coupling to baryons and cold dark matter (CDM) was originally performed in Refs. \cite{Amendola:1999er,Amendola:1999dr}. For some more recent works on interacting scalar field models see for example  \cite{Amendola:2003wa,Pettorino:2008ez,Gomez-Valent:2020mqn}. Furthermore, it is important to mention that 
a dark energy scalar field non-minimally coupled to baryons and dark matter can mediate additional long range forces on cosmological scales, but in order to ensure the concordance with the local gravitational experiments some screening mechanism is necessary to hide this extra gravitational interaction locally \cite{Khoury:2010xi,Babichev:2013usa} (for some alternative mechanism see Ref. \cite{Noller:2020lav}). Thus, as a very interesting way to extend the results obtained in the present paper and to proceed further in the general study of cosmological perturbations in these generalised scalar-torsion $f(T,\phi)$ gravity theories, one could allow an exchange of energy and momentum transfer between the scalar field and matter (cold dark matter and baryons) \cite{Kase:2019veo,Kase:2020hst}. These extensions lie beyond the scope of the present work and thus  are left for a separate project.

\appendix

\section{Appendix: Matrix coefficients }\label{Appendix1}
We list the matrix coefficients in Eq. \eqref{SOAction2}:
\bea
&& A_{11}=\frac{2 f_{,T} \left[3 (w+1) \bar{\rho} \left(2 T f_{,T}+P \dot{\phi}^2\right)+4 \frac{k^2}{a^2} P \dot{\phi}^2 f_{,T}\right]}{\frac{8}{3} \frac{k^2}{a^2} T f_{,T}^2-(w+1)\bar{\rho} \left(2 T f_{,T}+P \dot{\phi}^2\right)},\\
&& A_{12}=-\frac{2 \bar{\rho} f_{,T} \left(2 T f_{,T}+P \dot{\phi}^2\right)}{(w+1) \bar{\rho} \left(2 T f_{,T}+P \dot{\phi}^2\right)-\frac{8}{3} \frac{k^2}{a^2} T f_{,T}^2},\\
&& A_{22}= -\frac{\frac{4}{3} T \bar{\rho} f_{,T}^2}{(w+1) \left[(w+1) \bar{\rho} \left(2 T f_{,T}+P \dot{\phi}^2\right)-\frac{8}{3} \frac{k^2}{a^2} T f_{,T}^2\right]},\\
&& G_{11}= -2\Big[4 P \dot{\phi}^4 f_{,TT} \left(3 T P F_{,TT}^2 \left((w+1) \bar{\rho}-T f_{,T}\right)-\frac{3}{2} T P f_{,TT} f_{,T}^2+f_{,T\phi}^2 f_{,T}^2\right)+\nonumber\\
&& 24 H P \dot{\phi}^3 f_{,T} f_{,T\phi} f_{,TT} \left(f_{,TT} \left((w+1) \bar{\rho}-2 T f_{,T}\right)-f_{,T}^2\right)+\dot{\phi}^2 \Big(6 T P^3 \dot{\phi}^6 f_{,TT}^3+\nonumber\\
&& 4 (w+1) T \bar{\rho} P f_{,TT}^2 f_{,T}^2+6 (w+1) T \bar{\rho} P f_{,TT}^3 \left(\frac{2}{3} T f_{,T}+(w+1) \bar{\rho}\right)-4 f_{,T\phi}^2 f_{,T}^4\Big)+\nonumber\\
&& f_{,TT} f_{,T}^3 \left((w+1) \bar{\rho} P-8 T f_{,T\phi}^2\right)+24 H P^2 \dot{\phi}^5 f_{,T} f_{,T\phi} f_{,TT}^2+\nonumber\\
&& 2 (w+1) T \bar{\rho} f_{,T} f_{,TT} \left(2 T f_{,TT}+f_{,T}\right) \left(3 f_{,TT} \left(\frac{2}{3} T f_{,T}+(w+1) \bar{\rho}\right)+f_{,T}^2\right)\Big]/\nonumber\\
&& \Big[f_{,TT} \left(2 T f_{,TT}+f_{,T}\right)^2 \left(\frac{8}{3} k^2 T f_{,T}^2-(w+1) a^2 \bar{\rho} \left(2 T f_{,T}+P \dot{\phi}^2\right)\right)\Big],\\
&& G_{12}=\frac{2 \bar{\rho} f_{T} \left[f_{,T} \left((w+1) \bar{\rho}-4 H \dot{\phi} f_{,T,\phi}\right)-2T P \dot{\phi}^2 f_{,TT}\right]}{a^2\left(2 T f_{,TT}+f_{,T}\right) \left((w+1)\bar{\rho} \left(2 T f_{,T}+P \dot{\phi}^2\right)-\frac{8}{3} \frac{k^2}{a^2} T f_{,T}^2\right)},\\
&& G_{22}= -\frac{\frac{4}{3} w T \bar{\rho} f_{,T}^2}{(w+1)a^2 \left[(w+1) \bar{\rho} \left(2 T f_{,T}+P \dot{\phi}^2\right)-\frac{8}{3} \frac{k^2}{a^2} T f_{,T}^2\right]},\\
&& B_{22}=\frac{16 \frac{k^2}{a^2} f_{,T} \left[3 H f_{,TT} \left( P\dot{\phi}^2 \left(P \dot{\phi}^2+(w+1) \bar{\rho}\right)-\frac{8}{3} \frac{k^2}{a^2} T f_{,T}^2\right)+P \dot{\phi}^3 f_{,T} f_{,T\phi}-4 \frac{k^2}{a^2} H f_{,T}^3\right]}{\left(2 T f_{,TT}+f_{,T}\right) \left((w+1)\bar{\rho} \left(2 T f_{,T}+P \dot{\phi}^2\right)-\frac{8}{3} \frac{k^2}{a^2} T f_{,T}^2\right)},\\
&& B_{12}=\frac{16 k^2 H \bar{\rho} f_{,T}^2}{\frac{8}{3} k^2 T f_{,T}^2-(w+1) a^2 \bar{\rho} \left(2 T f_{,T}+P \dot{\phi}^2\right)},\\
&& B_{21}= \frac{4 \bar{\rho} \left[3 H f_{,TT} \left(a^2 P \dot{\phi}^2 \left(P \dot{\phi}^2+(w+1) \bar{\rho}\right)-\frac{8}{3} k^2 T f_{,T}^2\right)+a^2 P \dot{\phi}^3 f_{,T} f_{,T\phi}-4 k^2 H f_{,T}^3\right]}{\left(2 T f_{,TT}+f_{,T}\right) \left((w+1) a^2 \bar{\rho}  \left(2 T f_{,T}+P\dot{\phi}^2\right)-\frac{8}{3} k^2 T f_{,T}^2\right)},\\
&& B_{22}=\frac{4 a^2 H \bar{\rho}^2 f_{,T}}{\frac{8}{3} \frac{k^2}{a^2} T f_{,T}^2-(w+1) a^2 \bar{\rho} \left(2 T f_{,T}+P \dot{\phi}^2\right)},
\eea and \bea
&& M_{11}=\Bigg\{(w+1)\bar{\rho} \left(2 T \left(2 T f_{,TT}+f_{,T}\right)+P \dot{\phi}^2\right) \left(3 H f_{,TT} \left(P \dot{\phi}^2+(w+1) \bar{\rho}\right)+\dot{\phi} f_{,T} f_{,T\phi}\right)^2\nonumber\\
&& -\frac{8}{3} \frac{k^4}{a^4} T f_{,T}^2 f_{,TT} \left(2 T f_{,TT}+f_{,T}\right) \Bigg[2 T f_{,TT} \left(\frac{2}{3} T f_{,T}+2 P \dot{\phi}^2+(w+1) \bar{\rho}\right)-f_{,T} \Big((w+1) \bar{\rho}-\nonumber\\
&& 4 H \left(H f_{,T}+2 \dot{\phi} f_{,T\phi}\right)\Big)\Bigg]\Bigg\}/\Bigg[\frac{T}{3} f_{,TT} \left(2 T f_{,TT}+f_{,T}\right)^2 \Big((w+1)\bar{\rho} \left(2 T f_{,T}+P \dot{\phi}^2\right)-\nonumber\\
&& \frac{8}{3} \frac{k^2}{a^2} T f_{,T}^2\Big)\Bigg],\\
&& M_{12}=\frac{6 (w+1) H \bar{\rho}^2 \left[3 H f_{,TT} \left(P \dot{\phi}^2+(w+1) \bar{\rho}\right)+\dot{\phi} f_{,T} f_{,T\phi}\right]}{\left(2 T f_{,TT}+f_{,T}\right) \left((w+1) \bar{\rho} \left(2 T f_{,T}+P \dot{\phi}^2\right)-\frac{8}{3} \frac{k^2}{a^2} T f_{,T}^2\right)},\\
&& M_{22}=\frac{\bar{\rho}^2 \left(2 w T f_{,T}+w P \dot{\phi}^2+\bar{\rho}+w \bar{\rho}\right)}{2 (w+1) \bar{\rho} \left(2 T f_{,T}+P \dot{\phi}^2\right)-\frac{16}{3} \frac{k^2}{a^2} T f_{,T}^2}.
\eea The matrix coefficients of \eqref{L2_Cano_dia_I} are given by \cite{DeFelice:2016ucp}
\bea
&& K_{11}=A_{11}-\frac{A_{12}^2}{A_{22}}=\frac{P\dot{\phi}^2}{2 H^2},\\
&& K_{22}=k^2 A_{22}=-\frac{8 k^2 a^2 H^2 \rho f_{,T}^2}{(w+1) \left[(w+1) a^2 \rho \left(12 H^2 f_{,T}+P\dot{\phi}^2\right)-16 k^2 H^2 f_{,T}^2\right]},
\eea while the other matrix components are given by
\bea
&&Q_{12}=k \left[2\dot{A}_{12}-\frac{2 A_{12} \dot{A}_{22}}{A_{22}}-B_{12}+B_{21}\right],\nonumber\\
&& =\frac{2 k a^2 \rho\dot{\phi} \left[4 H f_{,T} \left(3 H P\dot{\phi}-f_{,\phi}\right)+P\dot{\phi} \left(P \dot{\phi}^2+(w+1) \rho\right)\right]}{(w+1) a^2 H \rho \left(12 H^2 f_{,T}+P\dot{\phi}^2\right)-16 k^2 H^3 f_{,T}^2},
\eea and
\bea
&& \mathcal{M}_{11}= \frac{1}{2 A_{22}^3}\Bigg\{A_{12} A_{22} \Big[A_{12} \left(\dot{B}_{22}+3 B_{22} H-2 k^2 G_{22}-2 M_{22}\right)+\nonumber\\
&& \dot{A}_{22} \left(B_{12}-B_{21}-4 \dot{A}_{12}\right)\Big]+\nonumber\\
&& A_{22}^2 \Big[A_{12} \left(-\dot{B}_{12}-3 H (B_{12}+B_{21})-\dot{B}_{21}+4 k^2 G_{12}+4 M_{12}\right)+\nonumber\\
&& \dot{A}_{12} \left(2 \dot{A}_{12}-B_{12}+B_{21}\right)\Big]+2 A_{12}^2 \dot{A}_{22}^2+\nonumber\\
&& A_{22}^3 \left(\dot{B}_{11}+3 B_{11} H-2 k^2 G_{11}-2 M_{11}\right)\Bigg\},
\eea

\bea
&& \mathcal{M}_{12}= \frac{1}{A_{22}^2}\Bigg\{k A_{22} \Big[B_{22} \dot{A}_{12}+2 A_{12} \left(k^2 G_{22}+M_{22}\right)-\nonumber\\
&& 2 A_{22} \left(k^2 G_{12}+M_{12}\right)\Big]-k A_{12} B_{22} \dot{A}_{22}\Bigg\},\\
&& \mathcal{M}_{22}=\frac{1}{2} k^2 \left(\dot{B}_{22}+3 B_{22} H-2 k^2 G_{22}-2 M_{22}\right).
\eea The matrix coefficients of \eqref{L2_Cano} are given by \cite{DeFelice:2016ucp}
\bea
&& \bar{B}(t,k)=\frac{A_{22} \left(2 \dot{A}_{12}-B_{12}+B_{21}\right)-2 A_{12} \dot{A}_{22}}{A_{22}^{3/2} \sqrt{A_{11}-\frac{A_{12}^2}{A_{22}}}},\\
&&  \bar{C}_{11}(t,k)=\Bigg[2 A_{12} A_{22} ^2 \Big(A_{22} \left(\dot{A}_{11} (-(B_{12}+B_{21}))-2 \dot{A}_{12} \left(\dot{A}_{11}+B_{11}\right)\right)+\nonumber\\
&& 2 A_{11} \left(-\dot{A}_{22} \left(2 \dot{A}_{12}+B_{21}\right)-B_{22} \dot{A}_{12}+2 A_{22} \left(k^2 G_{12}+M_{12}\right)\right)\Big)+\nonumber\\
&& 2 A_{12}^2 A_{22} \Big(A_{22} \left(\dot{A}_{22} \left(\dot{A}_{11}+B_{11}\right)+B_{22} \dot{A}_{11}-2 A_{11} \left(k^2 G_{22}+M_{22}\right)+2 B_{12} \dot{A}_{12}\right)+\nonumber\\
&& 2 A_{11} \dot{A}_{22} \left(\dot{A}_{22}+B_{22}\right)+2 A_{22}^2 \left(k^2 G_{11}+M_{11}\right)\Big)+A_{22}^3 \Big(A_{22} \dot{A}_{11} \left(\dot{A}_{11}+2 B_{11}\right)+\nonumber\\
&& 4 A_{11} \left(\dot{A}_{12} \left(\dot{A}_{12}+B_{21}\right)-A_{22}\left(k^2 G_{11}+M_{11}\right)\right)\Big)+2 A_{12}^3 A_{22} \Big(\dot{A}_{22} \left(2 \dot{A}_{12}-B_{12}+B_{21}\right)-\nonumber\\
&& 4 A_{22} \left(k^2 G_{12}+M_{12}\right)\Big)+A_{12}^4 \left(4 A_{22} \left(k^2 G_{22}+M_{22}\right)-\dot{A}_{22} \left(3 \dot{A}_{22}+2 B_{22}\right)\right)\Bigg]/\nonumber\\
&& \Bigg[4 A_{22}^2 \left(A_{12}^2-A_{11} A_{22}\right)^2\Bigg],\\
&& \bar{C}_{12}(t,k)= \Bigg[A_{22}^2 \Big(A_{22} B_{12} \dot{A}_{11}+A_{11} \Big(2 \dot{A}_{12} \left(\dot{A}_{22}+B_{22}\right)+B_{21} \dot{A}_{22}-\nonumber\\
&& 4 A_{22} \left(k^2 G_{12}+M_{12}\right)\Big)\Big)+A_{12} A_{22} \Big(A_{11} \Big(-3 B_{22} \dot{A}_{22}-2 \dot{A}_{22}^2+\nonumber\\
&& 4 A_{22} \left(k^2 G_{22}+M_{22}\right)\Big)-A_{22} \left(B_{22} \dot{A}_{11}+2 B_{12} \dot{A}_{12}\right)\Big)+\nonumber\\
&& A_{12}^2 A_{22} \left(\dot{A}_{22} \left(-2 \dot{A}_{12}+B_{12}-B_{21}\right)+4 A_{22} \left(k^2 G_{12}+M_{12}\right)\right)+\nonumber\\
&& 2 A_{12}^3 \left(\dot{A}_{22} \left(\dot{A}_{22}+B_{22}\right)-2 A_{22} \left(k^2 G_{22}+M_{22}\right)\right)\Bigg]/\nonumber\\
&& \Bigg[2 A_{22}^{5/2} \sqrt{A_{11}-\frac{A_{12}^2}{A_{22}}} \left(A_{11}A_{22}-A_{12}^2\right)\Bigg],\\
&& \bar{C}_{22}(t,k)=\frac{\dot{A}_{22} \left(\dot{A}_{22}+2 B_{22}\right)-4 A_{22} \left(k^2 G_{22}+M_{22}\right)}{4 A_{22}^2}.
\eea The expressions of these latter matrix components for the present model, which are obtained after replacing the matrix components of $A$, $B$, $G$, and $M$, are very complex and hence we do not show them explicitly.

\acknowledgments 

M. Gonzalez-Espinoza acknowledges support from PUCV. G. Otalora acknowldeges DI-VRIEA for financial support through Proyecto Postdoctorado $2020$ VRIEA-PUCV.


\bibliographystyle{spphys}       
\bibliography{bio}   

\begin{thebibliography}{100}
\providecommand{\url}[1]{{#1}}
\providecommand{\urlprefix}{URL }
\expandafter\ifx\csname urlstyle\endcsname\relax
  \providecommand{\doi}[1]{DOI \discretionary{}{}{}#1}\else
  \providecommand{\doi}{DOI \discretionary{}{}{}\begingroup
  \urlstyle{rm}\Url}\fi

\bibitem{Riess:1998cb}
A.G. Riess, et~al., Astron. J. \textbf{116}, 1009 (1998)

\bibitem{Perlmutter:1998np}
S.~Perlmutter, et~al., Astrophys. J. \textbf{517}, 565 (1999)

\bibitem{Aghanim:2018eyx}
N.~Aghanim, et~al., Astron. Astrophys. \textbf{641}, A6 (2020)

\bibitem{Ade:2015rim}
P.~Ade, et~al., Astron. Astrophys. \textbf{594}, A14 (2016)

\bibitem{Bull:2015stt}
P.~Bull, et~al., Phys. Dark Univ. \textbf{12}, 56 (2016)

\bibitem{Martin:2012bt}
J.~Martin, Comptes Rendus Physique \textbf{13}, 566 (2012)

\bibitem{Copeland:2006wr}
E.J. Copeland, M.~Sami, S.~Tsujikawa, Int. J. Mod. Phys. D \textbf{15}, 1753
  (2006)

\bibitem{amendola2010dark}
L.~Amendola, S.~Tsujikawa, \emph{Dark energy: theory and observations}
  (Cambridge University Press, 2010)

\bibitem{Riess:2011yx}
A.G. Riess, L.~Macri, S.~Casertano, H.~Lampeitl, H.C. Ferguson, A.V.
  Filippenko, S.W. Jha, W.~Li, R.~Chornock, Astrophys. J. \textbf{730}, 119
  (2011).
\newblock [Erratum: Astrophys.J. 732, 129 (2011)]

\bibitem{Riess:2016jrr}
A.G. Riess, et~al., Astrophys. J. \textbf{826}(1), 56 (2016)

\bibitem{Riess:2018byc}
A.G. Riess, et~al., Astrophys. J. \textbf{861}(2), 126 (2018)

\bibitem{DiValentino:2020zio}
E.~Di~Valentino, et~al., arXiv:2008.11284  (2020)

\bibitem{Hildebrandt:2016iqg}
H.~Hildebrandt, et~al., Mon. Not. Roy. Astron. Soc. \textbf{465}, 1454 (2017)

\bibitem{Kuijken:2015vca}
K.~Kuijken, et~al., Mon. Not. Roy. Astron. Soc. \textbf{454}(4), 3500 (2015)

\bibitem{Conti:2016gav}
I.~Fenech~Conti, R.~Herbonnet, H.~Hoekstra, J.~Merten, L.~Miller, M.~Viola,
  Mon. Not. Roy. Astron. Soc. \textbf{467}(2), 1627 (2017)

\bibitem{DiValentino:2018gcu}
E.~Di~Valentino, S.~Bridle, Symmetry \textbf{10}(11), 585 (2018)

\bibitem{DiValentino:2020vvd}
E.~Di~Valentino, et~al., arXiv:2008.11285  (2020)

\bibitem{Riess:2019cxk}
A.G. Riess, S.~Casertano, W.~Yuan, L.M. Macri, D.~Scolnic, Astrophys. J.
  \textbf{876}(1), 85 (2019)

\bibitem{Davari:2019tni}
Z.~Davari, V.~Marra, M.~Malekjani, Mon. Not. Roy. Astron. Soc. \textbf{491}(2),
  1920 (2020)

\bibitem{DiValentino:2015bja}
E.~Di~Valentino, A.~Melchiorri, J.~Silk, Phys. Rev. D \textbf{93}(2), 023513
  (2016)

\bibitem{Sola:2019jek}
J.~Sol\`a~Peracaula, A.~Gomez-Valent, J.~de~Cruz~P\'erez, C.~Moreno-Pulido,
  Astrophys. J. Lett. \textbf{886}(1), L6 (2019)

\bibitem{Sola:2020lba}
J.~Sola, A.~Gomez-Valent, J.d.C. Perez, C.~Moreno-Pulido, Class. Quant. Grav.
  \textbf{37}(24), 245003 (2020)

\bibitem{Joyce:2014kja}
A.~Joyce, B.~Jain, J.~Khoury, M.~Trodden, Phys. Rept. \textbf{568}, 1 (2015)

\bibitem{Koyama:2015vza}
K.~Koyama, Rept. Prog. Phys. \textbf{79}(4), 046902 (2016)

\bibitem{Wetterich:1987fm}
C.~Wetterich, Nucl. Phys. B \textbf{302}, 668 (1988)

\bibitem{Ratra:1987rm}
B.~Ratra, P.~Peebles, Phys. Rev. D \textbf{37}, 3406 (1988)

\bibitem{Carroll:1998zi}
S.M. Carroll, Phys. Rev. Lett. \textbf{81}, 3067 (1998)

\bibitem{Tsujikawa:2013fta}
S.~Tsujikawa, Class. Quant. Grav. \textbf{30}, 214003 (2013)

\bibitem{Chiba:1999ka}
T.~Chiba, T.~Okabe, M.~Yamaguchi, Phys. Rev. D \textbf{62}, 023511 (2000)

\bibitem{ArmendarizPicon:2000dh}
C.~Armendariz-Picon, V.F. Mukhanov, P.J. Steinhardt, Phys. Rev. Lett.
  \textbf{85}, 4438 (2000)

\bibitem{ArmendarizPicon:2000ah}
C.~Armendariz-Picon, V.F. Mukhanov, P.J. Steinhardt, Phys. Rev. D \textbf{63},
  103510 (2001)

\bibitem{Sen:2002nu}
A.~Sen, JHEP \textbf{04}, 048 (2002)

\bibitem{Sen:2002in}
A.~Sen, JHEP \textbf{07}, 065 (2002)

\bibitem{Linde:1982zj}
A.D. Linde, Phys. Lett. B \textbf{114}, 431 (1982)

\bibitem{Freedman:1974gs}
D.Z. Freedman, I.J. Muzinich, E.J. Weinberg, Annals Phys. \textbf{87}, 95
  (1974)

\bibitem{Freedman:1974ze}
D.Z. Freedman, E.J. Weinberg, Annals Phys. \textbf{87}, 354 (1974)

\bibitem{Birrell:1982ix}
N.D. Birrell, P.C.W. Davies, \emph{{Quantum Fields in Curved Space}} (Cambridge
  Univ. Press, Cambridge, UK, 1984)

\bibitem{Perrotta:1999am}
F.~Perrotta, C.~Baccigalupi, S.~Matarrese, Phys. Rev. D \textbf{61}, 023507
  (1999)

\bibitem{Sahni:1998at}
V.~Sahni, S.~Habib, Phys. Rev. Lett. \textbf{81}, 1766 (1998)

\bibitem{Chiba:1999wt}
T.~Chiba, Phys. Rev. D \textbf{60}, 083508 (1999)

\bibitem{Bartolo:1999sq}
N.~Bartolo, M.~Pietroni, Phys. Rev. D \textbf{61}, 023518 (2000)

\bibitem{Faraoni:2000wk}
V.~Faraoni, Phys. Rev. D \textbf{62}, 023504 (2000)

\bibitem{Hrycyna:2008gk}
O.~Hrycyna, M.~Szydlowski, JCAP \textbf{04}, 026 (2009)

\bibitem{Hrycyna:2007gd}
O.~Hrycyna, M.~Szydlowski, Phys. Rev. D \textbf{76}, 123510 (2007)

\bibitem{Sen:2008bg}
A.A. Sen, N.~Devi, Gen. Rel. Grav. \textbf{42}, 821 (2010)

\bibitem{deSouza:2008nj}
R.C. de~Souza, G.M. Kremer, Class. Quant. Grav. \textbf{26}, 135008 (2009)

\bibitem{Deffayet:2009wt}
C.~Deffayet, G.~Esposito-Farese, A.~Vikman, Phys. Rev. \textbf{D79}, 084003
  (2009)

\bibitem{DiValentino:2019jae}
E.~Di~Valentino, A.~Melchiorri, O.~Mena, S.~Vagnozzi, Phys. Rev. D
  \textbf{101}(6), 063502 (2020)

\bibitem{Einstein}
A.~Einstein, Sitz. Preuss. Akad. Wiss \textbf{217} (1928)

\bibitem{TranslationEinstein}
A.~Unzicker, T.~Case, arXiv:physics/0503046  (2005)

\bibitem{Early-papers1}
A.~Einstein, Math. Ann. \textbf{102}, 685 (1930)

\bibitem{Early-papers2}
A.~Einstein, Sitzungsber. Preuss. Akad. Wiss. Phys. Math. Kl. \textbf{401}
  (1930)

\bibitem{Early-papers3}
C.~Pellegrini, J.~Plebanski, Math.-Fys. Skr. Dan. Vid. Selskab \textbf{2}(2)
  (1962)

\bibitem{Early-papers4}
C.~M{\o}ller, K. Dan. Vidensk. Selsk., Mat.-Fys. Medd \textbf{39}(13), 1 (1978)

\bibitem{Early-papers5}
K.~Hayashi, T.~Nakano, Progress of Theoretical Physics \textbf{38}(2), 491
  (1967)

\bibitem{Early-papers6}
K.~Hayashi, T.~Shirafuji, Phys. Rev. D \textbf{19}(12), 3524 (1979)

\bibitem{JGPereira2}
J.G. Pereira, in \emph{Handbook of Spacetime}, ed. by A.~Ashtekar, V.~Petkov
  (Springer, 2014), pp. 197--212

\bibitem{AndradeGuillenPereira-00}
V.C. de~Andrade, L.C.T. Guillen, J.G. Pereira, Phys. Rev. Lett. \textbf{84},
  4533 (2000)

\bibitem{Arcos:2005ec}
H.I. Arcos, J.G. Pereira, Int. J. Mod. Phys. D \textbf{13}, 2193 (2004)

\bibitem{Pereira:2019woq}
J.G. Pereira, Y.N. Obukhov, Universe \textbf{5}(6), 139 (2019)

\bibitem{Aldrovandi-Pereira-book}
R.~Aldrovandi, J.G. Pereira, \emph{Teleparallel gravity: an introduction}, vol.
  173 (Springer Science \& Business Media, 2012)

\bibitem{Cai:2015emx}
Y.F. Cai, S.~Capozziello, M.~De~Laurentis, E.N. Saridakis, Rept. Prog. Phys.
  \textbf{79}(10), 106901 (2016)

\bibitem{Bahamonde:2017ize}
S.~Bahamonde, C.G. Böhmer, S.~Carloni, E.J. Copeland, W.~Fang, N.~Tamanini,
  Phys. Rept. \textbf{775-777}, 1 (2018)

\bibitem{Geng:2011aj}
C.Q. Geng, C.C. Lee, E.N. Saridakis, Y.P. Wu, Phys. Lett. B \textbf{704}, 384
  (2011)

\bibitem{Geng:2011ka}
C.Q. Geng, C.C. Lee, E.N. Saridakis, JCAP \textbf{1201}, 002 (2012)

\bibitem{Otalora:2013tba}
G.~Otalora, JCAP \textbf{1307}, 044 (2013)

\bibitem{Otalora:2013dsa}
G.~Otalora, Phys. Rev. D \textbf{88}, 063505 (2013)

\bibitem{Bengochea:2008gz}
G.R. Bengochea, R.~Ferraro, Phys. Rev. \textbf{D79}, 124019 (2009)

\bibitem{Linder:2010py}
E.V. Linder, Phys. Rev. \textbf{D81}, 127301 (2010)

\bibitem{Li:2011wu}
B.~Li, T.P. Sotiriou, J.D. Barrow, Phys. Rev. D \textbf{83}, 104017 (2011)

\bibitem{Harko:2014aja}
T.~Harko, F.S.N. Lobo, G.~Otalora, E.N. Saridakis, JCAP \textbf{12}, 021 (2014)

\bibitem{Harko:2014sja}
T.~Harko, F.S.N. Lobo, G.~Otalora, E.N. Saridakis, Phys. Rev. D \textbf{89},
  124036 (2014)

\bibitem{Carloni:2015lsa}
S.~Carloni, F.S. Lobo, G.~Otalora, E.N. Saridakis, Phys. Rev. D \textbf{93},
  024034 (2016)

\bibitem{Gonzalez-Espinoza:2018gyl}
M.~Gonzalez-Espinoza, G.~Otalora, J.~Saavedra, N.~Videla, Eur. Phys. J. C
  \textbf{78}(10), 799 (2018)

\bibitem{Nojiri:2004bi}
S.~Nojiri, S.D. Odintsov, Phys. Lett. B \textbf{599}, 137 (2004)

\bibitem{Allemandi:2005qs}
G.~Allemandi, A.~Borowiec, M.~Francaviglia, S.D. Odintsov, Phys. Rev. D
  \textbf{72}, 063505 (2005)

\bibitem{Nojiri:2006ri}
S.~Nojiri, S.D. Odintsov, eConf \textbf{C0602061}, 06 (2006)

\bibitem{Bertolami:2007gv}
O.~Bertolami, C.G. Boehmer, T.~Harko, F.S. Lobo, Phys. Rev. D \textbf{75},
  104016 (2007)

\bibitem{Harko:2008qz}
T.~Harko, Phys. Lett. B \textbf{669}, 376 (2008)

\bibitem{Harko:2010mv}
T.~Harko, F.S. Lobo, Eur. Phys. J. C \textbf{70}, 373 (2010)

\bibitem{Bertolami:2009ic}
O.~Bertolami, J.~Paramos, JCAP \textbf{03}, 009 (2010)

\bibitem{Bertolami:2013kca}
O.~Bertolami, P.~Fraz\~ao, J.~P\'aramos, JCAP \textbf{05}, 029 (2013)

\bibitem{Wang:2013fja}
J.~Wang, H.~Wang, Phys. Lett. B \textbf{724}, 5 (2013)

\bibitem{birrell1984quantum}
N.D. Birrell, N.D. Birrell, P.~Davies, P.~Davies, \emph{Quantum fields in
  curved space} (Cambridge university press, 1984)

\bibitem{Gonzalez-Espinoza:2020azh}
M.~Gonzalez-Espinoza, G.~Otalora, Phys. Lett. B \textbf{809}, 135696 (2020)

\bibitem{Gonzalez-Espinoza:2020jss}
M.~Gonzalez-Espinoza, G.~Otalora, arXiv:2011.08377  (2020)

\bibitem{Uzan:1999ch}
J.P. Uzan, Phys. Rev. D \textbf{59}, 123510 (1999)

\bibitem{Amendola:1999qq}
L.~Amendola, Phys. Rev. D \textbf{60}, 043501 (1999)

\bibitem{DeFelice:2016ucp}
A.~De~Felice, N.~Frusciante, G.~Papadomanolakis, JCAP \textbf{03}, 027 (2017)

\bibitem{Heisenberg:2016eld}
L.~Heisenberg, R.~Kase, S.~Tsujikawa, Phys. Lett. B \textbf{760}, 617 (2016)

\bibitem{Kase:2014cwa}
R.~Kase, S.~Tsujikawa, Int. J. Mod. Phys. D \textbf{23}(13), 1443008 (2015)

\bibitem{DeFelice:2011bh}
A.~De~Felice, S.~Tsujikawa, JCAP \textbf{02}, 007 (2012)

\bibitem{Sbisa:2014pzo}
F.~Sbis\`a, Eur. J. Phys. \textbf{36}, 015009 (2015)

\bibitem{Gergely:2014rna}
L.A. Gergely, S.~Tsujikawa, Phys. Rev. D \textbf{89}(6), 064059 (2014)

\bibitem{Gleyzes:2014qga}
J.~Gleyzes, D.~Langlois, F.~Piazza, F.~Vernizzi, JCAP \textbf{02}, 018 (2015)

\bibitem{Schutz:1977df}
B.F. Schutz, R.~Sorkin, Annals Phys. \textbf{107}, 1 (1977)

\bibitem{Brown:1992kc}
J.~Brown, Class. Quant. Grav. \textbf{10}, 1579 (1993)

\bibitem{Sotiriou:2010mv}
T.P. Sotiriou, B.~Li, J.D. Barrow, Phys. Rev. \textbf{D83}, 104030 (2011)

\bibitem{Li:2010cg}
B.~Li, T.P. Sotiriou, J.D. Barrow, Phys.\ Rev.\ D \textbf{83}, 064035 (2011)

\bibitem{Krssak:2015oua}
M.~Krššák, E.N. Saridakis, Class. Quant. Grav. \textbf{33}(11), 115009
  (2016)

\bibitem{Wu:2011kh}
Y.P. Wu, C.Q. Geng, Phys. Rev. D \textbf{86}, 104058 (2012)

\bibitem{DeFelice:2011uc}
A.~De~Felice, S.~Tsujikawa, Phys. Rev. D \textbf{84}, 083504 (2011)

\bibitem{Izumi:2012qj}
K.~Izumi, Y.C. Ong, JCAP \textbf{06}, 029 (2013)

\bibitem{Golovnev:2018wbh}
A.~Golovnev, T.~Koivisto, JCAP \textbf{1811}(11), 012 (2018)

\bibitem{Bluhm:2004ep}
R.~Bluhm, V.~Kostelecky, Phys. Rev. D \textbf{71}, 065008 (2005)

\bibitem{Bluhm:2007bd}
R.~Bluhm, S.H. Fung, V.A. Kostelecky, Phys. Rev. \textbf{D77}, 065020 (2008)

\bibitem{Wu:2016dkt}
Y.P. Wu, Phys. Lett. B \textbf{762}, 157 (2016)

\bibitem{Faddeev:1988qp}
L.D. Faddeev, R.~Jackiw, Phys. Rev. Lett. \textbf{60}, 1692 (1988)

\bibitem{Kase:2019veo}
R.~Kase, S.~Tsujikawa, Phys. Rev. D \textbf{101}(6), 063511 (2020)

\bibitem{Kase:2020hst}
R.~Kase, S.~Tsujikawa, JCAP \textbf{11}, 032 (2020)

\bibitem{DeFelice:2010gb}
A.~De~Felice, S.~Mukohyama, S.~Tsujikawa, Phys. Rev. D \textbf{82}, 023524
  (2010)

\bibitem{Arroja:2010wy}
F.~Arroja, M.~Sasaki, Phys. Rev. D \textbf{81}, 107301 (2010)

\bibitem{Giannakis:2005kr}
D.~Giannakis, W.~Hu, Phys. Rev. D \textbf{72}, 063502 (2005)

\bibitem{Garriga:1997wz}
J.~Garriga, X.~Montes, M.~Sasaki, T.~Tanaka, Nucl. Phys. B \textbf{513}, 343
  (1998).
\newblock [Erratum: Nucl.Phys.B 551, 511--511 (1999)]

\bibitem{Carroll:2003st}
S.M. Carroll, M.~Hoffman, M.~Trodden, Phys. Rev. D \textbf{68}, 023509 (2003)

\bibitem{Cline:2003gs}
J.M. Cline, S.~Jeon, G.D. Moore, Phys. Rev. D \textbf{70}, 043543 (2004)

\bibitem{Gumrukcuoglu:2016jbh}
A.E. G\"umr\"uk\c{c}\"uo\u{g}lu, S.~Mukohyama, T.P. Sotiriou, Phys. Rev. D
  \textbf{94}(6), 064001 (2016)

\bibitem{Frusciante:2016xoj}
N.~Frusciante, G.~Papadomanolakis, A.~Silvestri, JCAP \textbf{07}, 018 (2016)

\bibitem{Bean:2001wt}
R.~Bean, S.H. Hansen, A.~Melchiorri, Phys. Rev. D \textbf{64}, 103508 (2001)

\bibitem{Scherk:1980gq}
J.~Scherk, in \emph{{Europhysics Study Conference on Unification of the
  Fundamental Interactions}} (1980)

\bibitem{Sakstein:2017xjx}
J.~Sakstein, B.~Jain, Phys. Rev. Lett. \textbf{119}(25), 251303 (2017)

\bibitem{Bars:1986gt}
I.~Bars, M.~Visser, Phys. Rev. Lett. \textbf{57}, 25 (1986)

\bibitem{Hill:1988bu}
C.T. Hill, G.G. Ross, Nucl. Phys. B \textbf{311}, 253 (1988)

\bibitem{Fackler:1989nv}
O.~Fackler, J.~Tran Thanh~Van (eds.).
\newblock \emph{{Tests of fundamental laws in physics. Proceedings, 9th Moriond
  Workshop, 24th Rencontres de Moriond, Les Arcs, France, January 21-28, 1989}}
  (Ed. Frontieres, Gif-Sur-Yvette, 1989)

\bibitem{Damour:1992kf}
T.~Damour, K.~Nordtvedt, Phys. Rev. Lett. \textbf{70}, 2217 (1993)

\bibitem{Will:2014kxa}
C.M. Will, Living Rev. Rel. \textbf{17}, 4 (2014)

\bibitem{Ishak:2018his}
M.~Ishak, Living Rev. Rel. \textbf{22}(1), 1 (2019)

\bibitem{Burrage:2017qrf}
C.~Burrage, J.~Sakstein, Living Rev. Rel. \textbf{21}(1), 1 (2018)

\bibitem{LIGOScientific:2016aoc}
B.P. Abbott, et~al., Phys. Rev. Lett. \textbf{116}(6), 061102 (2016).
\newblock \doi{10.1103/PhysRevLett.116.061102}

\bibitem{LIGOScientific:2017ync}
B.P. Abbott, et~al., Astrophys. J. Lett. \textbf{848}(2), L12 (2017).
\newblock \doi{10.3847/2041-8213/aa91c9}

\bibitem{Frieman:1991zxc}
J.A. Frieman, B.A. Gradwohl, Phys. Rev. Lett. \textbf{67}, 2926 (1991)

\bibitem{Gradwohl:1992ue}
B.A. Gradwohl, J.A. Frieman, Astrophys. J. \textbf{398}, 407 (1992)

\bibitem{Gubser:2004du}
S.S. Gubser, P.J.E. Peebles, Phys. Rev. D \textbf{70}, 123511 (2004)

\bibitem{Gubser:2004uh}
S.S. Gubser, P.J.E. Peebles, Phys. Rev. D \textbf{70}, 123510 (2004)

\bibitem{Nusser:2004qu}
A.~Nusser, S.S. Gubser, P.J.E. Peebles, Phys. Rev. D \textbf{71}, 083505 (2005)

\bibitem{Kesden:2006zb}
M.~Kesden, M.~Kamionkowski, Phys. Rev. Lett. \textbf{97}, 131303 (2006)

\bibitem{Kesden:2006vz}
M.~Kesden, M.~Kamionkowski, Phys. Rev. D \textbf{74}, 083007 (2006)

\bibitem{Farrar:2006tb}
G.R. Farrar, R.A. Rosen, Phys. Rev. Lett. \textbf{98}, 171302 (2007)

\bibitem{Sealfon:2004gz}
C.~Sealfon, L.~Verde, R.~Jimenez, Phys. Rev. D \textbf{71}, 083004 (2005)

\bibitem{Bean:2008ac}
R.~Bean, E.E. Flanagan, I.~Laszlo, M.~Trodden, Phys. Rev. D \textbf{78}, 123514
  (2008)

\bibitem{Bai:2015vca}
Y.~Bai, J.~Salvado, B.A. Stefanek, JCAP \textbf{10}, 029 (2015)

\bibitem{Esteban:2021ozz}
I.~Esteban, J.~Salvado, JCAP \textbf{05}, 036 (2021)

\bibitem{Gonzalez-Espinoza:2019ajd}
M.~Gonzalez-Espinoza, G.~Otalora, N.~Videla, J.~Saavedra, JCAP \textbf{08}, 029
  (2019)

\bibitem{Baker:2017hug}
T.~Baker, E.~Bellini, P.G. Ferreira, M.~Lagos, J.~Noller, I.~Sawicki, Phys.
  Rev. Lett. \textbf{119}(25), 251301 (2017)

\bibitem{Flathmann:2019khc}
K.~Flathmann, M.~Hohmann, Phys. Rev. D \textbf{101}(2), 024005 (2020)

\bibitem{Chen:2014qsa}
Z.C. Chen, Y.~Wu, H.~Wei, Nucl. Phys. B \textbf{894}, 422 (2015)

\bibitem{Li:2013oef}
J.T. Li, Y.P. Wu, C.Q. Geng, Phys. Rev. D \textbf{89}(4), 044040 (2014)

\bibitem{Albuquerque:2018ymr}
I.S. Albuquerque, N.~Frusciante, N.J. Nunes, S.~Tsujikawa, Phys. Rev. D
  \textbf{98}(6), 064038 (2018)

\bibitem{Amendola:2006qi}
L.~Amendola, M.~Quartin, S.~Tsujikawa, I.~Waga, Phys. Rev. D \textbf{74},
  023525 (2006)

\bibitem{Gomes:2013ema}
A.~Gomes, L.~Amendola, JCAP \textbf{03}, 041 (2014)

\bibitem{Tino:2020nla}
G.M. Tino, L.~Cacciapuoti, S.~Capozziello, G.~Lambiase, F.~Sorrentino, Prog.
  Part. Nucl. Phys. \textbf{112}, 103772 (2020)

\bibitem{Wetterich:1994bg}
C.~Wetterich, Astron. Astrophys. \textbf{301}, 321 (1995)

\bibitem{Damour:1990tw}
T.~Damour, G.W. Gibbons, C.~Gundlach, Phys. Rev. Lett. \textbf{64}, 123 (1990)

\bibitem{Amendola:1999er}
L.~Amendola, Phys. Rev. D \textbf{62}, 043511 (2000)

\bibitem{Amendola:1999dr}
L.~Amendola, Mon. Not. Roy. Astron. Soc. \textbf{312}, 521 (2000)

\bibitem{Amendola:2003wa}
L.~Amendola, Phys. Rev. D \textbf{69}, 103524 (2004)

\bibitem{Pettorino:2008ez}
V.~Pettorino, C.~Baccigalupi, Phys. Rev. D \textbf{77}, 103003 (2008)

\bibitem{Gomez-Valent:2020mqn}
A.~G\'omez-Valent, V.~Pettorino, L.~Amendola, Phys. Rev. D \textbf{101}(12),
  123513 (2020)

\bibitem{Khoury:2010xi}
J.~Khoury, arXiv:1011.5909 [astro-ph.CO]  (2010)

\bibitem{Babichev:2013usa}
E.~Babichev, C.~Deffayet, Class. Quant. Grav. \textbf{30}, 184001 (2013)

\bibitem{Noller:2020lav}
J.~Noller, L.~Santoni, E.~Trincherini, L.G. Trombetta, JCAP \textbf{01}, 045
  (2021)

\end{thebibliography}



\end{document}